\documentclass[nofootinbib,preprint,preprintnumbers,a4paper,10pt]{revtex4}
\usepackage{multirow}
\usepackage{textcomp}
\usepackage{amsmath,graphicx,color,epsfig}

\usepackage{footnote}
\usepackage{ulem}
\usepackage{booktabs}
\usepackage{array}
\usepackage{amssymb}
\usepackage{subfigure}
\usepackage{longtable}
\usepackage{verbatim}
\usepackage{amsfonts}
\usepackage{hyperref}
\usepackage{cancel}

\begin{document}

\title{Topological diagrams of doubly charmed baryon decays in the $SU(3)_F$ limit}

\author{Si-Hong Liu$^{1}$}
\author{Ying-Xin Lai$^{1,2}$}
\author{Di Wang$^{2}$}\email{wangdi@hunnu.edu.cn}
\address{%
$^1$Hunan University of Science and Technology, Xiangtan 411201, China\\
$^2$Hunan Normal University, Changsha 410081, China
}

\begin{abstract}
The doubly charmed baryon was first observed by LHCb via the non-leptonic decay $\Xi_{cc}^{++}\to \Lambda^+_cK^-\pi^+\pi^+$ in 2017.
Subsequently, ongoing efforts have been made to identify other doubly charmed baryons.
However, there is no systematic analysis of the topological decomposition for non-leptonic decays of doubly charmed baryons.
In this work, we study the topological amplitudes of doubly charmed baryon decays in the $SU(3)_F$ limit.
Tree- and penguin-induced topological diagrams for the $\mathcal{B}_{cc}\to \mathcal{B}_c M$ and $\mathcal{B}_{cc}\to \mathcal{B} D$ decays are presented.
The linear relations between the topological amplitudes and the $SU(3)$ irreducible amplitudes are derived through tensor contraction and $SU(3)$ decomposition.
The magnitude pattern of the topological diagrams is analyzed in the rescattering dynamics and the large $N_c$ expansion.
In addition, some amplitude relations are derived to test the K\"orner-Pati-Woo theorem in the isospin limit.

\end{abstract}

\maketitle

\section{Introduction}

Doubly heavy baryons provide a unique system for studying heavy-to-light baryonic transitions and non-perturbative strong interactions.
The LHCb Collaboration discovered the doubly charmed baryon $\Xi_{cc}^{++}$ via the decay $\Xi_{cc}^{++}\to \Lambda^+_cK^-\pi^+\pi^+$ in 2017 \cite{Aaij:2017ueg}.
Subsequently, the Large Hadron Collider beauty (LHCb) Collaboration measured the lifetime of $\Xi_{cc}^{++}$ and observed several decay modes \cite{LHCb:2025shu,LHCb:2019ybf,LHCb:2022rpd,Aaij:2018wzf,Aaij:2018gfl,LHCb:2019qed}.
Other doubly heavy baryons have also been studied by the LHCb Collaboration, but no signals have been observed to date \cite{LHCb:2022fbu,LHCb:2021eaf,LHCb:2021rkb,LHCb:2021xba,LHCb:2020iko,LHCb:2019gqy}.
From a theoretical perspective, the semileptonic weak decays of doubly charmed baryons have been studied extensively, while non-leptonic weak decays have received relatively little attention.
The heavy-to-light non-leptonic decays involve large non-perturbative effects.
Obtaining reliable predictions in the factorization analysis is challenging due to the lack of knowledge about low-energy inputs and the complicated hard-scattering kernels.
The discovery of $\Xi_{cc}^{++}$ benefited from meson-baryon rescattering dynamics \cite{Yu:2017zst}, which have been intensively studied for doubly charmed baryon decays \cite{Li:2020qrh,Hu:2024uia,Jiang:2018oak,Han:2021azw}.

Apart from model calculations, it is useful to study non-leptonic decays of doubly charmed baryons based on $SU(3)_F$ symmetry \cite{Li:2021rfj,Shi:2017dto,Wang:2017azm}.
Topological diagrams are a popular tool for analyzing heavy hadron weak decays based on flavor symmetry, as they intuitively describe the internal dynamics of hadron decays at the quark level.
However, to date there has been no systematic study of topological decomposition of doubly charmed baryon decays.
In this work, we study the topological amplitudes of doubly charmed baryon two-body non-leptonic decays in the $SU(3)_F$ limit, following the theoretical framework constructed in Refs.~\cite{Wang:2025bdl,Wang:2024ztg,Wang:2024nxb}.
Tree- and penguin-induced diagrams of $\mathcal{B}_{cc}\to \mathcal{B}_{c\overline 3} M$, $\mathcal{B}_{cc}\to \mathcal{B}_{c6} M$, $\mathcal{B}_{cc}\to \mathcal{B}_{8} D$, and $\mathcal{B}_{cc}\to \mathcal{B}_{10} D$ are presented in detail.
The linear relations of the topological amplitudes and the $SU(3)$ irreducible amplitudes are derived through tensor contraction and $SU(3)$ decomposition.
The magnitude pattern of topological diagrams is analyzed within the frameworks of rescattering dynamics and large $N_c$ expansion.
The K\"orner-Pati-Woo theorem \cite{Pati:1970fg,Korner:1970xq} is found to be inconsistent with rescattering dynamics \cite{Li:2020qrh,Hu:2024uia,Jiang:2018oak,Han:2021azw,Jia:2024pyb,Wang:2025khg}.
Several equations are derived to test the K\"orner-Pati-Woo theorem in the isospin limit.

The rest of this paper is organized as follows.
In Sect. \ref{to}, we present the topological amplitudes for the $\mathcal{B}_{cc}\to \mathcal{B}_c M$ and $\mathcal{B}_{cc}\to \mathcal{B} D$ decays.
The phenomenological analysis for the doubly charmed baryon decays is performed in
Sect. \ref{pa}, and a brief summary is provided in Sect. \ref{summary}.
The $U$-spin sum rules for $\mathcal{B}_{cc}\to \mathcal{B}_{8}D$ decays are listed in Appendix~\ref{U}.

\section{Topological amplitudes and linear relations}\label{to}

\subsection{Decays into a charmed baryon and a light meson}\label{BM}

In this section, we present the topological diagrams of charmed baryon decays into a charmed baryon and a light meson.
The effective Hamiltonian in charm quark decay in the Standard Model (SM) can be written as \cite{Buchalla:1995vs}
 \begin{equation}\label{hsm}
 \mathcal H_{\rm eff}={G_F\over \sqrt 2}
 \left[\sum_{q=d,s}V_{cq_1}^*V_{uq_2}\left(\sum_{q=1}^2C_i(\mu)\mathcal{O}_i(\mu)\right)
 -V_{cb}^*V_{ub}\left(\sum_{i=3}^6C_i(\mu)\mathcal{O}_i(\mu)+C_{8g}(\mu)\mathcal{O}_{8g}(\mu)\right)\right],
 \end{equation}
 where $G_F$ is the Fermi coupling constant and $C_{i}$ is the Wilson coefficient of operator $\mathcal{O}_i$.
The magnetic-penguin contributions can be included in the Wilson coefficients for the penguin operators following the substitutions
\cite{Beneke:1999br}
\begin{eqnarray}
C_{3,5}(\mu)\to& C_{3,5}(\mu) + \frac{\alpha_s(\mu)}{8\pi N_c}
\frac{2m_c^2}{\langle l^2\rangle}C_{8g}^{\rm eff}(\mu),\qquad
C_{4,6}(\mu)\to& C_{4,6}(\mu) - \frac{\alpha_s(\mu)}{8\pi }
\frac{2m_c^2}{\langle l^2\rangle}C_{8g}^{\rm eff}(\mu),\label{mag}
\end{eqnarray}
with the effective Wilson coefficient $C_{8g}^{\rm eff}=C_{8g}+C_5$.
In the $SU(3)$ framework, the weak Hamiltonian of charm decay can be written as \cite{Wang:2020gmn}
 \begin{equation}\label{h}
 \mathcal H_{\rm eff}= \sum_p \sum_{i,j,k=1}^3 (H^{(p)})_{ij}^{k}\mathcal{O}_{ij}^{(p)k},
 \end{equation}
in which
\begin{equation}
\mathcal{O}_{ij}^{(p)k} = \frac{G_F}{\sqrt{2}} \sum_{\rm color} \sum_{\rm current}C_p(\overline q_iq_k)(\overline q_jc),
\end{equation}
and $p=0$ and $p=1$ denote the tree and penguin operators, respectively.
The coefficients $(H^{(p)})_{ij}^k$ can be obtained from the mapping $(\bar uq_1)(\bar q_2c)\rightarrow V^*_{cq_2}V_{uq_1}$ for current-current operators and $(\bar qq)(\bar uc)\rightarrow -V^*_{cb}V_{ub}$ for penguin operators.
The nonzero $(H^{(0)})_{ij}^k$ induced by tree operators are
\begin{align}\label{ckm1}
 &(H^{(0)})_{13}^2 = V_{cs}^*V_{ud},  \qquad (H^{(0)})^{2}_{12}=V_{cd}^*V_{ud},\qquad (H^{(0)})^{3}_{13}= V_{cs}^*V_{us}, \qquad (H^{(0)})^{3}_{12}=V_{cd}^*V_{us}.
\end{align}
 The nonzero $(H^{(1)})_{ij}^k$ induced by penguin operators are
\begin{align}\label{ckm2}
 &(H^{(1)})_{11}^1 = -V_{cb}^*V_{ub}, \qquad (H^{(1)})_{21}^2=-V_{cb}^*V_{ub}, \qquad (H^{(1)})_{31}^3=-V_{cb}^*V_{ub}.
\end{align}

The doubly charmed triplet baryon is
\begin{align}
  \mathcal{B}_{cc}  = (\Xi_{cc}^{++},\,\,\Xi_{cc}^{+},\,\, \Omega_{cc}^{+} ).
\end{align}
The charmed anti-triplet baryon is
\begin{eqnarray}
 \mathcal{B}_{c\overline 3}=  \left( \begin{array}{ccc}
   0   & \Lambda_c^+  & \Xi_c^+ \\
    -\Lambda_c^+ &   0   & \Xi_c^0 \\
    -\Xi_c^+ & -\Xi_c^0 & 0 \\
  \end{array}\right).
\end{eqnarray}
The charmed sextet baryon is
\begin{eqnarray}
 \mathcal{B}_{c6}=  \left( \begin{array}{ccc}
   \Sigma_c^{++}   &  \frac{1}{\sqrt{2}}\Sigma_c^{+}  & \frac{1}{\sqrt{2}}\Xi_c^{*+} \\
   \frac{1}{\sqrt{2}}\Sigma_c^{+} &   \Sigma_c^{0}   & \frac{1}{\sqrt{2}}\Xi_c^{*0} \\
    \frac{1}{\sqrt{2}}\Xi_c^{*+} & \frac{1}{\sqrt{2}}\Xi_c^{*0} & \Omega_c^0 \\
  \end{array}\right).
\end{eqnarray}
The light pseudoscalar nonet meson is
\begin{eqnarray}
 M=  \left( \begin{array}{ccc}
   \frac{1}{\sqrt 2} \pi^0+  \frac{1}{\sqrt 6} \eta_8    & \pi^+  & K^+ \\
    \pi^- &   - \frac{1}{\sqrt 2} \pi^0+ \frac{1}{\sqrt 6} \eta_8   & K^0 \\
    K^- & \overline K^0 & -\sqrt{2/3}\eta_8 \\
  \end{array}\right) +  \frac{1}{\sqrt 3} \left( \begin{array}{ccc}
   \eta_1    & 0  & 0 \\
    0 &  \eta_1   & 0 \\
   0 & 0 & \eta_1 \\
  \end{array}\right).
\end{eqnarray}
The amplitude of doubly charmed baryon decays into a charmed anti-triplet baryon and a light meson can be constructed by contracting all indices of the four-quark Hamiltonian with the initial/final states,
\begin{align}\label{am1}
  \mathcal{A}(\mathcal{B}_{cc}\to \mathcal{B}_{c\overline3}M) &=  A_1(\mathcal{B}_{cc})_iH^i_{jk}M^j_l\mathcal{B}_{c\overline3}^{kl}+  A_2(\mathcal{B}_{cc})_iH^i_{jk}M^l_l\mathcal{B}_{c\overline3}^{jk}+ A_3(\mathcal{B}_{cc})_iH^j_{lk}M^i_j\mathcal{B}_{c\overline 3}^{lk}+ A_4(\mathcal{B}_{cc})_iH^l_{jk}M^k_l\mathcal{B}_{c\overline3}^{ij}\nonumber\\
   & +A_5(\mathcal{B}_{cc})_iH^l_{kj}M^k_l\mathcal{B}_{c\overline3}^{ij}+  A_6(\mathcal{B}_{cc})_iH^i_{kj}M^j_l\mathcal{B}_{c\overline3}^{kl}+ A_7(\mathcal{B}_{cc})_iH^l_{jl}M^j_k\mathcal{B}_{c\overline3}^{ik}+ A_8(\mathcal{B}_{cc})_iH^l_{jl}M^i_k\mathcal{B}_{c\overline3}^{jk}\nonumber\\
   &+A_9(\mathcal{B}_{cc})_iH^l_{jl}M^k_k \mathcal{B}_{c\overline3}^{ij}
   + A_{10}(\mathcal{B}_{cc})_iH^l_{lj}M^j_k\mathcal{B}_{c\overline3}^{ik}+ A_{11}(\mathcal{B}_{cc})_iH^l_{lj}M^i_k \mathcal{B}_{c\overline3}^{jk}\nonumber\\&
   +A_{12}(\mathcal{B}_{cc})_iH^l_{lj}M^k_k \mathcal{B}_{c\overline3}^{ij}.
\end{align}
Similarly, the amplitude of $\mathcal{B}_{cc}\to \mathcal{B}_{c6}M$ decay is constructed as
\begin{align}\label{am2}
  \mathcal{A}_{\rm eff}(\mathcal{B}_{cc}\to \mathcal{B}_{c6}M)& =  A^\prime_1(\mathcal{B}_{cc})_iH^i_{jk}M^j_l\mathcal{B}_{c6}^{kl}+  A^\prime_2(\mathcal{B}_{cc})_iH^i_{jk}M^l_l\mathcal{B}_{c6}^{jk}+ A^\prime_3(\mathcal{B}_{cc})_iH^j_{lk}M^i_j\mathcal{B}_{c6}^{lk}+ A^\prime_4(\mathcal{B}_{cc})_iH^l_{jk}M^k_l\mathcal{B}_{c6}^{ij}\nonumber\\
   & +A^\prime_5(\mathcal{B}_{cc})_iH^l_{kj}M^k_l\mathcal{B}_{c6}^{ij}+  A^\prime_6(\mathcal{B}_{cc})_iH^i_{kj}M^j_l\mathcal{B}_{c6}^{kl}+ A^\prime_7(\mathcal{B}_{cc})_iH^l_{jl}M^j_k\mathcal{B}_{c6}^{ik}+ A^\prime_8(\mathcal{B}_{cc})_iH^l_{jl}M^i_k\mathcal{B}_{c6}^{jk}\nonumber\\
   &+A^\prime_9(\mathcal{B}_{cc})_iH^l_{jl}M^k_k\mathcal{B}_{c6}^{ij}
   + A^\prime_{10}(\mathcal{B}_{cc})_iH^l_{lj}M^j_k\mathcal{B}_{c6}^{ik}+ A^\prime_{11}(\mathcal{B}_{cc})_iH^l_{lj}M^i_k\mathcal{B}_{c6}^{jk}
  \nonumber\\
   & +A^\prime_{12}(\mathcal{B}_{cc})_iH^l_{lj}M^k_k\mathcal{B}_{c6}^{ij}.
\end{align}
The completeness of Eqs.~\eqref{am1} and \eqref{am2} can be verified through permutation.
The sum of decay amplitudes contributing to $\mathcal{B}_{cc}\to \mathcal{B}_{c\overline 3}M$ and $\mathcal{B}_{cc}\to \mathcal{B}_{c6}M$ modes is $N_{\overline 3}+N_6=A^4_4=24$.

\begin{figure}
  \centering
  \includegraphics[width=10cm]{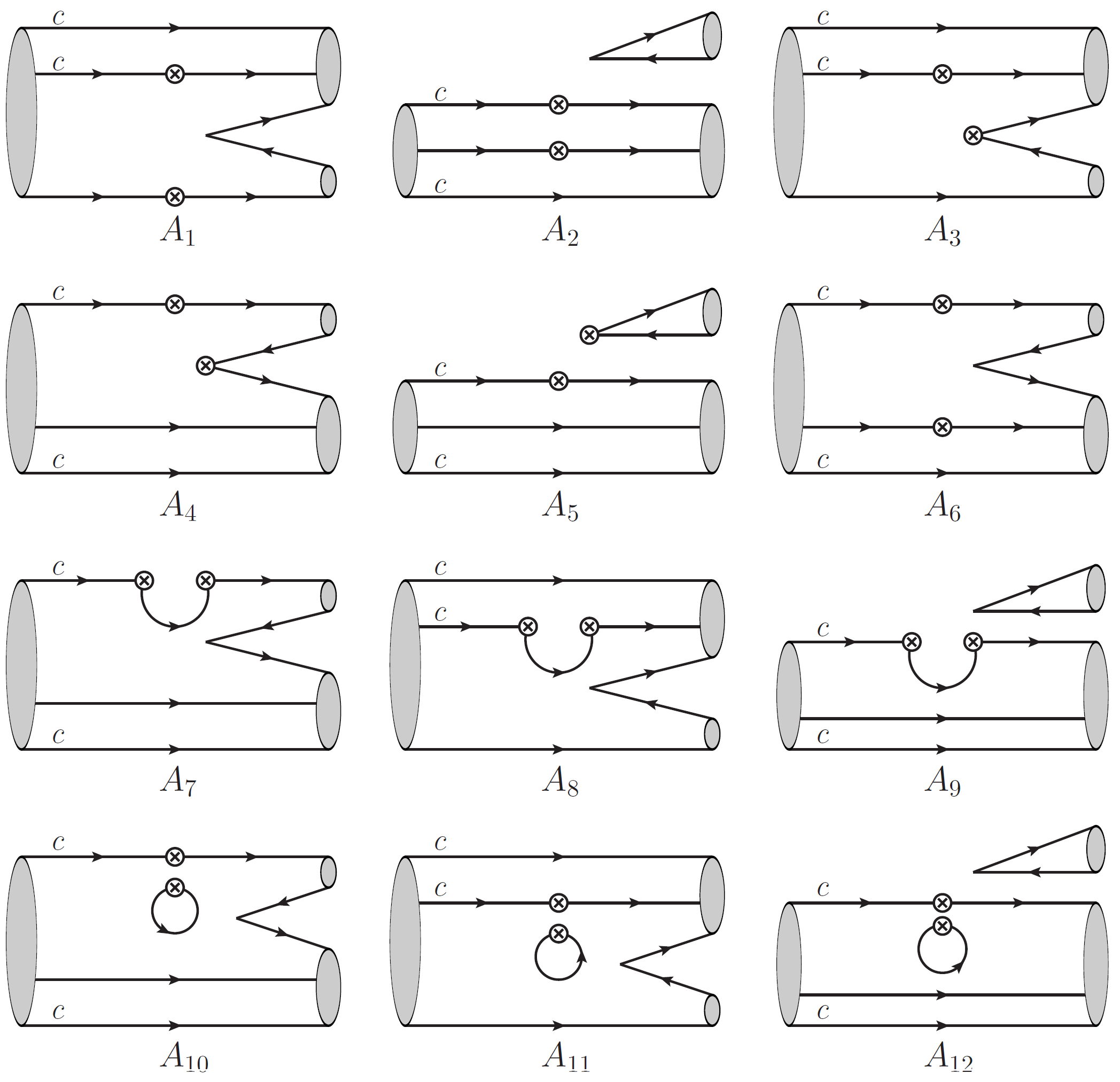}
  \caption{Topological diagrams of the $\Xi_{cc}$ and $\Omega_{cc}$ decays into a charmed baryon and a light meson.}\label{top1}
\end{figure}

\begin{table*}
\caption{Topological amplitudes of the $\mathcal{B}_{cc}\to \mathcal{B}_{c\overline 3}M$ decays.}\label{amp1}
 \small
\begin{tabular}{|c|c|c|c|}
\hline\hline
 Channel & \qquad\qquad\qquad Amplitude \qquad\qquad\qquad & Channel &Amplitude\\\hline
 $\Xi_{cc}^{++}\to \Xi_c^+\pi^+$ & $\lambda_1(A_3+A_5)$ &   $\Xi_{cc}^{++}\to \Lambda_c^+K^+$ & $\lambda_2(A_3+A_5)$  \\\hline
  $\Xi_{cc}^{+}\to \Lambda_c^+\overline K^0$ & $\lambda_1(-A_4+A_6)$ &    $\Xi_{cc}^{+}\to \Lambda_c^+K^0$ & $\lambda_2(A_3-A_4)$ \\\hline
  $\Xi_{cc}^{+}\to \Xi_c^+\pi^0$ & $\frac{1}{\sqrt{2}}\lambda_1(-A_1-A_3)$ &    $\Omega_{cc}^{+}\to \Lambda_c^+\pi^0$ & $\frac{1}{\sqrt{2}}\lambda_2(-A_1-A_6)$\\\hline
  $\Xi_{cc}^{+}\to \Xi_c^+\eta_8$ & $\frac{1}{\sqrt{6}}\lambda_1(-A_1+A_3-2A_6)$ &  $\Omega_{cc}^{+}\to \Lambda_c^+\eta_8$ & $\frac{1}{\sqrt{6}}\lambda_2(-A_1-2A_3+A_6)$\\\hline
  $\Xi_{cc}^{+}\to \Xi_c^+\eta_1$ & ~~~~~~~~$\frac{1}{\sqrt{3}}\lambda_1(-A_1+3A_2+A_3+A_6)$~~~~~~~~ &    $\Omega_{cc}^{+}\to \Lambda_c^+\eta_1$ & $\frac{1}{\sqrt{3}}\lambda_2(-A_1+3A_2+A_3+A_6)$\\\hline
  $\Xi_{cc}^{+}\to \Xi_c^0\pi^+$ & $\lambda_1(-A_1+A_5)$ & $\Omega_{cc}^{+}\to \Xi_c^+K^0$ & $\lambda_2(-A_4+A_6)$  \\\hline
  $\Omega_{cc}^{+}\to \Xi_c^+\overline K^0$ & $\lambda_1(A_3-A_4)$ & $\Omega_{cc}^{+}\to \Xi_c^0K^+$ & $\lambda_2(A_1-A_5)$ \\\toprule[1.2pt]
 $\Xi_{cc}^{++}\to \Lambda_c^+\pi^+$ & \multicolumn{3}{c|}{$\lambda_d(A_3+A_5+A_7+A_8)+\lambda_s(A_7+A_8)-\lambda_b(A_1^P-A_3^P+A_4^P+A_6^P+A_7^P+A_8^P+3A_{10}^P+3A_{11}^P)$}   \\\hline
 $\Xi_{cc}^{++}\to \Xi_c^+K^+$ &
  \multicolumn{3}{c|}{$\lambda_d(A_7+A_8)+\lambda_s(A_3+A_5+A_7+A_8)-\lambda_b(A_1^P-A_3^P+A_4^P+A_6^P+A_7^P+A_8^P+3A_{10}^P+3A_{11}^P)$}  \\\hline
 $\Xi_{cc}^{+}\to \Lambda_c^+\pi^0$ &
 \multicolumn{3}{c|}{ $\frac{1}{\sqrt{2}}\lambda_d(-A_1-A_3+A_4-A_6-A_7-A_8)-\frac{1}{\sqrt{2}}\lambda_s(A_7+A_8)$}\\
 & \multicolumn{3}{c|}{$-\frac{1}{\sqrt{2}}\lambda_b(-A_1^P+A_3^P-A_4^P-A_6^P-A_7^P-A_8^P-3A_{10}^P-3A_{11}^P)$}   \\\hline
 $\Xi_{cc}^{+}\to \Lambda_c^+\eta_8$ &
  \multicolumn{3}{c|}{$\frac{1}{\sqrt{6}}\lambda_d(-A_1+A_3-A_4+A_6-A_7+A_8)+\frac{1}{\sqrt{6}}\lambda_s(2A_4-A_7+A_8)$}\\
 & \multicolumn{3}{c|}{$-\frac{1}{\sqrt{6}}\lambda_b(A_1^P-A_3^P-A_4^P-A_6^P-A_7^P+A_8^P-3A_{10}^P+3A_{11}^P)$}    \\\hline
 $\Xi_{cc}^{+}\to \Lambda_c^+\eta_1$ &  \multicolumn{3}{c|}{$\frac{1}{\sqrt{3}}\lambda_d(-A_1+3A_2+A_3-A_4+A_6-A_7+A_8-3A_9)+\frac{1}{\sqrt{3}}\lambda_s(-A_4-A_7+A_8-3A_9)$}\\
 & \multicolumn{3}{c|}{$-\frac{1}{\sqrt{6}}\lambda_b(A_1^P-3A_2^P-A_3^P-A_4^P-3A_5^P-A_6^P-A_7^P+A_8^P-3A^P_9-3A_{10}^P+3A_{11}^P-9A_{12}^P)$}    \\\hline
 $\Xi_{cc}^{+}\to \Xi_c^+K^0$ & \multicolumn{3}{c|}{$\lambda_d(A_6+A_8)+\lambda_s(A_3+A_8)-\lambda_b(A_1^P-A_3^P+A_8^P+3A_{11}^P)$} \\\hline
 $\Xi_{cc}^{+}\to \Xi_c^0K^+$ &  \multicolumn{3}{c|}{$\lambda_d(A_1+A_7)+\lambda_s(A_5+A_7)-\lambda_b(A_4^P+A_6^P+A_7^P+3A_{10}^P)$} \\\hline
 $\Omega_{cc}^{+}\to \Lambda_c^+\overline K^0$ &  \multicolumn{3}{c|}{ $\lambda_d(A_3+A_8)+\lambda_s(A_6+A_8)-\lambda_b(A_1^P-A_3^P+A_8^P+3A_{11}^P)$} \\\hline
 $\Omega_{cc}^{+}\to \Xi_c^+ \pi^0$ & \multicolumn{3}{c|}{  $\frac{1}{\sqrt{2}}\lambda_d(A_4-A_7)+\frac{1}{\sqrt{2}}\lambda_s(-A_1-A_7)
 +\frac{1}{\sqrt{2}}\lambda_b(A_4^P+A_6^P+A_7^P+3A_{10}^P)$} \\\hline
 $\Omega_{cc}^{+}\to \Xi_c^+ \eta_8$ &  \multicolumn{3}{c|}{$\frac{1}{\sqrt{6}}\lambda_d(-A_4-A_7-2A_8)+\frac{1}{\sqrt{6}}\lambda_s(-A_1-2A_3+2A_4-2A_6-A_7-2A_8)$}\\
 & \multicolumn{3}{c|}{$-\frac{1}{\sqrt{6}}\lambda_b(-2A_1^P+2A_3^P-A_4^P-A_6^P-A_7^P-2A_8^P-3A_{10}^P-6A_{11}^P)$}    \\\hline
 $\Omega_{cc}^{+}\to \Xi_c^+ \eta_1$ &  \multicolumn{3}{c|}{$\frac{1}{2\sqrt{3}}\lambda_d(-A_4-A_7+A_8-3A_9)+\frac{1}{\sqrt{3}}\lambda_s(- A_1+3A_2+A_3-A_4+A_6-A_7+A_8-3A_9)$}\\
 & \multicolumn{3}{c|}{$-\frac{1}{\sqrt{6}}\lambda_b(A_1^P-3A_2^P-A_3^P-A_4^P-3A^P_5-A_6^P-A_7^P+A_8^P-3A^P_9-3A_{10}^P+3A_{11}^P-9A_{12}^P)$}    \\\hline
 $\Omega_{cc}^{+}\to \Xi_c^0\pi^+$ &   \multicolumn{3}{c|}{$\lambda_d(-A_5-A_7)+\lambda_s(-A_1-A_7)
 +\lambda_b(A_4^P+A_6^P+A_7^P+3A_{10}^P)$} \\\hline
  \hline
\end{tabular}
\end{table*}

\begin{table*}
\caption{Topological amplitudes of the $\mathcal{B}_{cc}\to \mathcal{B}_{c6}M$ decays.}\label{amp2}
 \small
\begin{tabular}{|c|c|c|c|}
\hline\hline
 Channel & \qquad\qquad\qquad Amplitude \qquad\qquad\qquad & Channel &Amplitude\\\hline
 $\Xi_{cc}^{++}\to \Sigma_c^{++}\overline K^0$ & $\lambda_1A^\prime_4$ &$\Xi_{cc}^{++}\to \Sigma_c^{++} K^0$ & $\lambda_2A^\prime_4$ \\\hline
 $\Xi_{cc}^{++}\to \Xi_c^{*+}\pi^+$ & $\frac{1}{\sqrt{2}}\lambda_1(A^\prime_3+A^\prime_5)$&$\Xi_{cc}^{++}\to \Sigma_c^{+}K^+$ & $\frac{1}{\sqrt{2}}\lambda_2(A^\prime_3+A^\prime_5)$ \\\hline
 $\Xi_{cc}^{+}\to \Sigma_c^{++}K^-$ &$\lambda_1A^\prime_6$ &$\Xi_{cc}^{+}\to \Sigma_c^{+}K^0$ &$\frac{1}{\sqrt{2}}\lambda_2(A^\prime_3+A^\prime_4)$  \\\hline
 $\Xi_{cc}^{+}\to \Sigma_c^{+}\overline K^0$ &$\frac{1}{\sqrt{2}}\lambda_1(A^\prime_4+A^\prime_6)$ &$\Xi_{cc}^{+}\to \Sigma_c^{0} K^+$ & $\lambda_2A^\prime_5$ \\\hline
 $\Xi_{cc}^{+}\to \Xi_c^{*+}\pi^0$ &$\frac{1}{2}\lambda_1(A^\prime_1-A^\prime_3)$ & $\Omega_{cc}^{+}\to \Sigma_c^{++}\pi^-$& $\lambda_2A^\prime_6$ \\\hline
 $\Xi_{cc}^{+}\to \Xi_c^{*+}\eta_8$ &$\frac{1}{2\sqrt{3}}\lambda_1(A^\prime_1+A^\prime_3-2A^\prime_6)$ &$\Omega_{cc}^{+}\to \Sigma_c^{+}\pi^0$ & $\frac{1}{2}\lambda_2(A^\prime_1-A^\prime_6)$ \\\hline
 $\Xi_{cc}^{+}\to \Xi_c^{*+}\eta_1$ & $\frac{1}{\sqrt{6}}\lambda_1(A^\prime_1+3A^\prime_2+A^\prime_3+A^\prime_6)$&$\Omega_{cc}^{+}\to \Sigma_c^{+}\eta_8$ & $\frac{1}{2\sqrt{3}}\lambda_2(A^\prime_1-2A^\prime_3+A^\prime_6)$ \\\hline
 $\Xi_{cc}^{+}\to \Xi_c^{*0}\pi^+$ & $\frac{1}{\sqrt{2}}\lambda_1(A^\prime_1+A^\prime_5)$&$\Omega_{cc}^{+}\to \Sigma_c^{+}\eta_1$ & $\frac{1}{\sqrt{6}}\lambda_2(A^\prime_1+3A^\prime_2+A^\prime_3+A^\prime_6)$ \\\hline
 $\Xi_{cc}^{+}\to \Omega_c^{0}K^+$ & $\lambda_1 A^\prime_1$& $\Omega_{cc}^{+}\to \Sigma_c^{0}\pi^+$& $\lambda_2A^\prime_1$ \\\hline
 $\Omega_{cc}^{+}\to \Xi_c^{*+}\overline K^0$ &$\frac{1}{\sqrt{2}}\lambda_1(A^\prime_3+A^\prime_4)$ &$\Omega_{cc}^{+}\to \Xi_c^{*+}K^0$ &$\frac{1}{\sqrt{2}}\lambda_2(A^\prime_4+A^\prime_6)$  \\\hline
 $\Omega_{cc}^{+}\to \Omega_c^{0} \pi^+$ & $\lambda_1 A^\prime_5$&$\Omega_{cc}^{+}\to \Xi_c^{*0}K^+$ & $\frac{1}{\sqrt{2}}\lambda_2(A^\prime_1+A^\prime_5)$ \\\toprule[1.2pt]
$\Xi_{cc}^{++}\to \Sigma_c^{++}\pi^0$ &  \multicolumn{3}{c|}{ $\frac{1}{\sqrt{2}}\lambda_d(-A^\prime_4+A^\prime_7+A^\prime_8)
+\frac{1}{\sqrt{2}}\lambda_s(A^\prime_7+A^\prime_8)$ } \\ &
 \multicolumn{3}{c|}{$-\frac{1}{\sqrt{2}}\lambda_b(A^{\prime P}_1+A^{\prime P}_3+A^{\prime P}_4+A^{\prime P}_6+A^{\prime P}_7+A^{\prime P}_8+3A^{\prime P}_{10}+3A^{\prime P}_{11})$ } \\\hline
$\Xi_{cc}^{++}\to \Sigma_c^{++}\eta_8$ &   \multicolumn{3}{c|}{ $\frac{1}{\sqrt{6}}\lambda_d(A^\prime_4+A^\prime_7+A^\prime_8)
+\frac{1}{\sqrt{6}}\lambda_s(-2A^\prime_4+A^\prime_7+A^\prime_8)$ } \\ &
 \multicolumn{3}{c|}{$-\frac{1}{\sqrt{6}}\lambda_b(A^{\prime P}_1+A^{\prime P}_3+A^{\prime P}_4+A^{\prime P}_6+A^{\prime P}_7+A^{\prime P}_8+3A^{\prime P}_{10}+3A^{\prime P}_{11})$ } \\\hline
$\Xi_{cc}^{++}\to \Sigma_c^{++}\eta_1$ &  \multicolumn{3}{c|}{ $\frac{1}{\sqrt{3}}\lambda_d(A^\prime_4+A^\prime_7+A^\prime_8+3A^\prime_9)
+\frac{1}{\sqrt{3}}\lambda_s(A^\prime_4+A^\prime_7+A^\prime_8+3A^\prime_9)$ } \\ &
 \multicolumn{3}{c|}{$-\frac{1}{\sqrt{3}}\lambda_b(A^{\prime P}_1+3A^{\prime P}_2+A^{\prime P}_3+A^{\prime P}_4+3A^{\prime P}_5+A^{\prime P}_6+A^{\prime P}_7+A^{\prime P}_8+3A^{\prime P}_{9}+3A^{\prime P}_{10}+3A^{\prime P}_{11}+9A^{\prime P}_{12})$ } \\\hline
$\Xi_{cc}^{++}\to \Sigma_c^{+}\pi^+$ &  \multicolumn{3}{c|}{ $\frac{1}{\sqrt{2}}\lambda_d(A^\prime_3+A^\prime_5+A^\prime_7+A^\prime_8)
+\frac{1}{\sqrt{2}}\lambda_s(A^\prime_7+A^\prime_8)$ } \\ &
 \multicolumn{3}{c|}{$-\frac{1}{\sqrt{2}}\lambda_b(A^{\prime P}_1+A^{\prime P}_3+A^{\prime P}_4+A^{\prime P}_6+A^{\prime P}_7+A^{\prime P}_8+3A^{\prime P}_{10}+3A^{\prime P}_{11})$ } \\\hline
$\Xi_{cc}^{++}\to \Xi_c^{*+}K^+$ &  \multicolumn{3}{c|}{ $\frac{1}{\sqrt{2}}\lambda_d(A^\prime_7+A^\prime_8)
+\frac{1}{\sqrt{2}}\lambda_s(A^\prime_3+A^\prime_5+A^\prime_7+A^\prime_8)$ } \\ &
 \multicolumn{3}{c|}{$-\frac{1}{\sqrt{2}}\lambda_b(A^{\prime P}_1+A^{\prime P}_3+A^{\prime P}_4+A^{\prime P}_6+A^{\prime P}_7+A^{\prime P}_8+3A^{\prime P}_{10}+3A^{\prime P}_{11})$ } \\\hline
$\Xi_{cc}^{+}\to \Sigma_c^{++}\pi^-$ &  \multicolumn{3}{c|}{ $\lambda_d(A^\prime_6+A^\prime_8)+\lambda_sA^\prime_8-\lambda_b(A^{\prime P}_1+A^{\prime P}_3+A^{\prime P}_8+3A^{\prime P}_{11})$ } \\\hline
$\Xi_{cc}^{+}\to \Sigma_c^{+}\pi^0$ & \multicolumn{3}{c|}{ $\frac{1}{2}\lambda_d(A^\prime_1-A^\prime_3-A^\prime_4-A^\prime_6+A^\prime_7-A^\prime_8)
+\frac{1}{2}\lambda_s(A^\prime_7-A^\prime_8)$ } \\ &
 \multicolumn{3}{c|}{$-\frac{1}{2}\lambda_b(-A^{\prime P}_1-A^{\prime P}_3+A^{\prime P}_4+A^{\prime P}_6+A^{\prime P}_7-A^{\prime P}_8+3A^{\prime P}_{10}-3A^{\prime P}_{11})$ } \\\hline
$\Xi_{cc}^{+}\to \Sigma_c^{+}\eta_8$ &  \multicolumn{3}{c|}{ $\frac{1}{2\sqrt{3}}\lambda_d(A^\prime_1+A^\prime_3+A^\prime_4+A^\prime_6+A^\prime_7+A^\prime_8)
+\frac{1}{2\sqrt{3}}\lambda_s(-2A^\prime_4+A^\prime_7+A^\prime_8)$ } \\ &
 \multicolumn{3}{c|}{$-\frac{1}{2\sqrt{3}}\lambda_b(A^{\prime P}_1+A^{\prime P}_3+A^{\prime P}_4+A^{\prime P}_6+A^{\prime P}_7+A^{\prime P}_8+3A^{\prime P}_{10}+3A^{\prime P}_{11})$ } \\\hline
$\Xi_{cc}^{+}\to \Sigma_c^{+}\eta_1$ &  \multicolumn{3}{c|}{ $\frac{1}{\sqrt{6}}\lambda_d(A^\prime_1+3A^\prime_2+A^\prime_3+A^\prime_4+A^\prime_6
+A^\prime_7+A^\prime_8+3A^\prime_9)+\frac{1}{2\sqrt{3}}\lambda_s(A^\prime_4+A^\prime_7
+A^\prime_8+3A^\prime_9)$ } \\ &
 \multicolumn{3}{c|}{$-\frac{1}{\sqrt{6}}\lambda_b(A^{\prime P}_1+3A^{\prime P}_2+A^{\prime P}_3+A^{\prime P}_4+3A^{\prime P}_5+A^{\prime P}_6+A^{\prime P}_7+A^{\prime P}_8+3A^{\prime P}_9+3A^{\prime P}_{10}+3A^{\prime P}_{11}+9A^{\prime P}_{12})$ } \\\hline
$\Xi_{cc}^{+}\to \Sigma_c^{0}\pi^+$ &  \multicolumn{3}{c|}{ $\lambda_d(A^\prime_1+A^\prime_5+A^\prime_7)+\lambda_sA^\prime_7-\lambda_b(A^{\prime P}_4+A^{\prime P}_6+A^{\prime P}_7+3A^{\prime P}_{10})$ } \\\hline
$\Xi_{cc}^{+}\to \Xi_c^{*+}K^0$ &  \multicolumn{3}{c|}{ $\frac{1}{\sqrt{2}}\lambda_d(A^\prime_6+A^\prime_8)
+\frac{1}{\sqrt{2}}\lambda_s(A^\prime_3+A^\prime_8)-\frac{1}{\sqrt{2}}\lambda_b(A^{\prime P}_1+A^{\prime P}_3+A^{\prime P}_8+3A^{\prime P}_{11})$ } \\\hline
$\Xi_{cc}^{+}\to \Xi_c^{*0}K^+$ &  \multicolumn{3}{c|}{ $\frac{1}{\sqrt{2}}\lambda_d(A^\prime_1+A^\prime_7)
+\frac{1}{\sqrt{2}}\lambda_s(A^\prime_5+A^\prime_7)-\frac{1}{\sqrt{2}}\lambda_b(A^{\prime P}_4+A^{\prime P}_6+A^{\prime P}_7+3A^{\prime P}_{10})$ } \\\hline
$\Omega_{cc}^{+}\to \Sigma_c^{++}K^-$ &  \multicolumn{3}{c|}{ $\lambda_dA^\prime_8+\lambda_s(A^\prime_6+A^\prime_8)-\lambda_b(A^{\prime P}_1+A^{\prime P}_3+A^{\prime P}_8+3A^{\prime P}_{11})$ } \\\hline
$\Omega_{cc}^{+}\to \Sigma_c^{+}\overline K^0$ & \multicolumn{3}{c|}{ $\frac{1}{\sqrt{2}}\lambda_d(A^\prime_3+A^\prime_8)
+\frac{1}{\sqrt{2}}\lambda_s(A^\prime_6+A^\prime_8)-\frac{1}{\sqrt{2}}\lambda_b(A^{\prime P}_1+A^{\prime P}_3+A^{\prime P}_8+3A^{\prime P}_{11})$ } \\\hline
$\Omega_{cc}^{+}\to \Xi_c^{*+}\pi^0$ & \multicolumn{3}{c|}{ $\frac{1}{2}\lambda_d(-A^\prime_4+A^\prime_7)+\frac{1}{2}\lambda_s(A^\prime_1+A^\prime_7)
-\frac{1}{2}\lambda_b(A^{\prime P}_4+A^{\prime P}_6+A^{\prime P}_7+3A^{\prime P}_{10})$ } \\\hline
$\Omega_{cc}^{+}\to \Xi_c^{*+}\eta_8$ &  \multicolumn{3}{c|}{ $\frac{1}{2\sqrt{3}}\lambda_d(A^\prime_4+A^\prime_7-2A^\prime_8)
+\frac{1}{2\sqrt{3}}\lambda_s(A^\prime_1-2A^\prime_3-2A^\prime_4-2A^\prime_6
+A^\prime_7-2A^\prime_8)$ } \\ &
 \multicolumn{3}{c|}{$-\frac{1}{2\sqrt{3}}\lambda_b(-2A^{\prime P}_1-2A^{\prime P}_3+A^{\prime P}_4+A^{\prime P}_6+A^{\prime P}_7-2A^{\prime P}_8+3A^{\prime P}_{10}-6A^{\prime P}_{11})$ } \\\hline
$\Omega_{cc}^{+}\to \Xi_c^{*+}\eta_1$ &  \multicolumn{3}{c|}{ $\frac{1}{\sqrt{6}}\lambda_d(A^\prime_4+A^\prime_7+A^\prime_8+3A^\prime_9)
+\frac{1}{\sqrt{6}}\lambda_s(A^\prime_1+3A^\prime_2+A^\prime_3+A^\prime_4+A^\prime_6
+A^\prime_7+A^\prime_8+3A^\prime_9)$ } \\ &
 \multicolumn{3}{c|}{$-\frac{1}{\sqrt{6}}\lambda_b(A^{\prime P}_1+3A^{\prime P}_2+A^{\prime P}_3+A^{\prime P}_4+3A^{\prime P}_5+A^{\prime P}_6+A^{\prime P}_7+A^{\prime P}_8+3A^{\prime P}_{9}+3A^{\prime P}_{10}+3A^{\prime P}_{11}+9A^{\prime P}_{12})$ } \\\hline
$\Omega_{cc}^{+}\to \Xi_c^{*0}\pi^+$ & \multicolumn{3}{c|}{ $\frac{1}{\sqrt{2}}\lambda_d(A^\prime_5+A^\prime_7)
+\frac{1}{\sqrt{2}}\lambda_s(A^\prime_1+A^\prime_7)-\frac{1}{\sqrt{2}}\lambda_b(A^{\prime P}_4+A^{\prime P}_6+A^{\prime P}_7+3A^{\prime P}_{10})$ } \\\hline
$\Omega_{cc}^{+}\to \Omega_c^{0}K^+$ &  \multicolumn{3}{c|}{ $\lambda_dA^\prime_7+\lambda_s(A^\prime_1+A^\prime_5+A^\prime_7)-\lambda_b(A^{\prime P}_4+A^{\prime P}_6+A^{\prime P}_7+3A^{\prime P}_{10})$ } \\\hline
  \hline
\end{tabular}
\end{table*}

If index contraction is understood as quark flow, each term in Eqs.~\eqref{am1} and \eqref{am2} represents a topological diagram with the following rules.
\begin{itemize}
  \item In the tensor $H_{ij}^k$, the index $j$ represents the quark $q_j$ produced at the vertex that connects to the $c$ quark line, while the indices $i$ and $k$ denote the quark $q_i$ and the antiquark $\overline q_k$ produced at the other vertex.
  \item The contraction of the final-state meson or baryon with the four-quark Hamiltonian indicates that the quark produced at one vertex enters the final-state meson or baryon.
  \item  The contraction between the initial and final states indicates that the light quark in the charmed baryon enters the final state as a spectator quark.
\item  The contraction between two indices in $H_{ij}^k$ denotes a quark loop. There are two distinct ways to contract two indices in $H_{ij}^k$: $H_{il}^l$ represents the quark loop connecting the two vertices in the topological diagram, and $H_{lj}^l$ represents the quark loop induced by a single vertex in the topological diagram.
\end{itemize}
In this work, we classify topological diagrams as tree-induced or penguin-induced diagrams according to which operators, tree or penguin, are inserted into the weak vertices, irrespective of whether the diagram includes a quark loop or not.

The amplitudes of $\mathcal{B}_{cc}\to \mathcal{B}_{c\overline3}M$ and $\mathcal{B}_{cc}\to \mathcal{B}_{c6}M$ decays are listed in Tables~\ref{amp1} and \ref{amp2}, respectively.
The penguin-induced amplitudes are labeled with a superscript "$P$" to distinguish them from the tree-induced amplitudes.
Note that there are nine tree-induced amplitudes, $A_{1}\sim A_9$ ($A^\prime_{1}\sim A^\prime_9$), and 12 penguin-induced amplitudes, $A^P_{1}\sim A^P_{12}$ ($A^{\prime P}_{1}\sim A^{\prime P}_{12}$), contributing to the $\mathcal{B}_{cc}\to \mathcal{B}_{c\overline3}M$ ($\mathcal{B}_{cc}\to \mathcal{B}_{c6}M$) decays.

In the $SU(3)$ framework, the weak operator $\mathcal{O}^{k}_{ij}$ is decomposed as
\begin{align}\label{su}
  \mathcal{O}^k_{ij}= &\mathcal{O}(15)^k_{ij}+\epsilon_{ijl}\mathcal{O}(\overline 6)^{lk}+\delta_j^k\Big(\frac{3}{8}\mathcal{O}( 3)_i-\frac{1}{8}\mathcal{O}(3^\prime)_i\Big)+
  \delta_i^k\Big(\frac{3}{8}\mathcal{O}( 3^\prime)_j-\frac{1}{8}\mathcal{O}( 3)_j\Big).
\end{align}
The $SU(3)$ irreducible amplitude of $\mathcal{B}_{cc}\to \mathcal{B}_{c\overline3}M$ decay is
\begin{align}\label{su3}
  \mathcal{A}^{IR}(\mathcal{B}_{cc}\to \mathcal{B}_{c\overline3}M) &=  a_1(\mathcal{B}_{cc})_iH(15)^i_{jk}M^j_l\mathcal{B}_{c\overline3}^{kl}+ a_2(\mathcal{B}_{cc})_iH(15)^l_{jk}M^k_l\mathcal{B}_{c\overline3}^{ij}
  +a_3(\mathcal{B}_{cc})_iH(\overline 6)^i_{jk}M^j_l \mathcal{B}_{c\overline3}^{kl}\nonumber\\&
  +a_4(\mathcal{B}_{cc})_iH(\overline 6)^l_{jk}M^k_l \mathcal{B}_{c\overline3}^{ij}+ a_5(\mathcal{B}_{cc})_iH(\overline 6)^j_{lk}M^i_j\mathcal{B}_{c\overline 3}^{lk}+  a_6(\mathcal{B}_{cc})_iH(\overline 6)^i_{jk}M^l_l \mathcal{B}_{c\overline3}^{jk}\nonumber\\
   &
   + a_7(\mathcal{B}_{cc})_iH(3)_{j}M^j_k\mathcal{B}_{c\overline3}^{ik}+ a_8(\mathcal{B}_{cc})_iH(3)_{j}M^i_k\mathcal{B}_{c\overline3}^{jk}
   +a_9(\mathcal{B}_{cc})_iH(3)_{j}M^k_k\mathcal{B}_{c\overline3}^{ij}
   \nonumber\\&+ a_{10}(\mathcal{B}_{cc})_iH(3^\prime)_{j}M^j_k \mathcal{B}_{c\overline3}^{ik}+ a_{11}(\mathcal{B}_{cc})_iH(3^\prime)_{j}M^i_k\mathcal{B}_{c\overline3}^{jk}
   +a_{12}(\mathcal{B}_{cc})_iH(3^\prime)_{j}M^k_k\mathcal{B}_{c\overline3}^{ij}.
\end{align}
According to Eq.~\eqref{su}, the relations between the $SU(3)$ irreducible amplitudes and the topological amplitudes in $\mathcal{B}_{cc}\to \mathcal{B}_{c\overline3}M$ decays are derived as
\begin{align}\label{sol}
   & a_1 = A_1 + A_6,\qquad a_2 = A_4 + A_5,  \qquad a_3 = A_1 - A_6,  \qquad a_4 = A_4 - A_5, \qquad a_5 = A_3 \qquad a_6 = A_2, \nonumber\\
   &  a_7 = \frac{3}{8}A_1-  \frac{1}{8}A_4+ \frac{3}{8}A_5 -  \frac{1}{8}A_6 + A_7,\quad a_8 = -\frac{1}{8}A_1  +  \frac{1}{2}A_3 +\frac{3}{8}A_6  + A_8,\quad  a_9 = -\frac{1}{2}A_2 +  \frac{3}{8}A_4- \frac{1}{8}A_5 + A_9,\nonumber\\
    &  a_{10} = -\frac{1}{8}A_1 +  \frac{3}{8}A_4- \frac{1}{8}A_5+\frac{3}{8}A_6 + A_{10},\quad a_{11} = \frac{3}{8}A_1 - \frac{1}{2}A_3 -\frac{1}{8}A_6  + A_{11},\nonumber\\ &  a_{12} = \frac{1}{2}A_2 -  \frac{1}{8}A_4+ \frac{3}{8}A_5 + A_{12}.
\end{align}
With Levi-Civita symbol, the four terms constructed by $H(\overline 6)_{ij}^k$ in Eq.~\eqref{su3} can be written as
\begin{align}\label{xx}
a_3(\mathcal{B}_{cc})_iH(\overline 6)^i_{jk}M^j_l\mathcal{B}_{c\overline3}^{kl}&= -a_3(\mathcal{B}_{cc})_iH(\overline 6)^{ij}M^l_l({\mathcal{B}_{c\overline3}})_{j}+a_3(\mathcal{B}_{cc})_iH(\overline 6)^{ij}M^k_j({\mathcal{B}_{c\overline3}})_{k},\nonumber\\
a_4(\mathcal{B}_{cc})_iH(\overline 6)^l_{jk}M^k_l\mathcal{B}_{c\overline3}^{ij}&=-a_4(\mathcal{B}_{cc})_iH(\overline 6)^{jk}M^i_j({\mathcal{B}_{c\overline3}})_{k}+a_4(\mathcal{B}_{cc})_iH(\overline 6)^{ij}M^k_j({\mathcal{B}_{c\overline3}})_{k},\nonumber\\ a_5(\mathcal{B}_{cc})_i\mathcal{H}(\overline 6)^j_{lk}M^i_j\mathcal{B}_{c\overline 3}^{lk}& =2a_5(\mathcal{B}_{cc})_iH(\overline 6)^{jk}M^i_j({\mathcal{B}_{c\overline3}})_{k},\nonumber\\ a_6(\mathcal{B}_{cc})_iH(\overline 6)^i_{jk}M^l_l\mathcal{B}_{c\overline3}^{jk}&=2a_6(\mathcal{B}_{cc})_iH(\overline 6)^{ij}M^l_l({\mathcal{B}_{c\overline3}})_{j}.
\end{align}
We can define three new parameters to replace $a_3\sim a_6$,
\begin{align}\label{redef}
a^\prime_3 = a_3 + a_4,\qquad a^\prime_4 = a_4 - 2a_5, \qquad a^\prime_5 = a_3 - 2a_6.
\end{align}

The linear correlation of decay amplitudes contributing to the $\mathcal{B}_{cc}\to \mathcal{B}_{c\overline 3}M$ modes in the SM is beyond the tensor analysis in Eq.~\eqref{xx}.
The nonzero coefficients induced by tree operators in the $SU(3)$ irreducible representations are given by
\begin{align}\label{ckm3}
 &  H^{(0)}( \overline6)^{22}=-\frac{V_{cs}^*V_{ud}}{2},\qquad H^{(0)}( \overline 6)^{23}=\frac{V_{cd}^*V_{ud}-V_{cs}^*V_{us}}{4},  \qquad H^{(0)}( \overline 6)^{33}=  \frac{V_{cd}^*V_{us}}{2},\nonumber \\
   &  H^{(0)}(15)^{1}_{11}=-\frac{V_{cd}^*V_{ud}+V_{cs}^*V_{us}}{4}, \qquad H^{(0)}(15)^{2}_{13}= \frac{V_{cs}^*V_{ud}}{2},  \qquad  H^{(0)}(15)^{3}_{12}=\frac{V_{cd}^*V_{us}}{2},\nonumber \\
 &  H^{(0)}(15)^{2}_{12}= \frac{3V_{cd}^*V_{ud}-V_{cs}^*V_{us}}{8},\qquad H^{(0)}(15)^{3}_{13}=\frac{3V_{cs}^*V_{us}-V_{cd}^*V_{ud}}{8},\qquad H^{(0)}( 3)_1=V_{cd}^*V_{ud}+V_{cs}^*V_{us},
\end{align}
and the nonzero coefficients induced by penguin operators in the $SU(3)$ irreducible representations are given by
\begin{align}\label{ckm4}
 H^{(1)}( 3)_1=-V_{cb}^*V_{ub}, \qquad H^{(1)}( 3^\prime)_1=-3V_{cb}^*V_{ub}.
\end{align}
Note that the three-dimensional representation $3^\prime$ is absent in Eq.~\eqref{ckm3}, and the fifteen- and six-dimensional representations are absent in Eq.~\eqref{ckm4}.
This indicates that only the tree-induced $SU(3)$ irreducible amplitudes $a_1\sim a_{9}$ and the penguin-induced $SU(3)$ irreducible amplitudes $a_7^P\sim a_{12}^P$ contribute to the $\mathcal{B}_{cc}\to \mathcal{B}_{c\overline 3}M$ decays.
Moreover, the coefficient matrices of the three-dimensional representations, including $H^{(0)}(3)$,  $H^{(1)}(3)$, and $H^{(1)}(3^\prime)$, contain only the first component.
Due to the unitarity of the Cabibbo-Kobayashi-Maskawa (CKM) matrix, we have $H^{(0)}(3)_1 = V_{cd}^*V_{ud}+V_{cs}^*V_{us} = -V_{cb}^*V_{ub}$, and thus
$H^{(0)}(3)_1:H^{(1)}(3)_1:H^{(1)}(3^\prime)_1 = -V_{cb}^*V_{ub}:-V_{cb}^*V_{ub}:-3V_{cb}^*V_{ub}$.
The $SU(3)$ irreducible amplitudes involving $H^{(0)}(3)$,  $H^{(1)}(3)$, and $H^{(1)}(3^\prime)$ appear in the following fixed combinations:
\begin{align}\label{x1}
  a_7^{T+P} &= a_7 + a^P_7 + 3a^P_{10},\qquad a_8^{T+P} = a_8 + a^P_8 + 3a^P_{11}, \qquad
  a_9^{T+P} = a_9 + a^P_9 + 3a^P_{12}.
\end{align}
According to Eq.~\eqref{x1}, the penguin-induced $SU(3)$ irreducible amplitudes $a_7^P\sim a_{12}^P$ always appear together with tree-induced $SU(3)$ irreducible amplitudes $a_{7}$, $a_8$, and $a_9$.
Substituting Eq.~\eqref{sol} into to Eq.~\eqref{x1}, we conclude that the penguin-induced diagrams $A_{1}^P\sim A_{12}^P$ appear together with the tree-induced diagrams $A_{7}$, $A_8$, and $A_9$ in the following fixed combinations:
\begin{align}
  A_7^{T+P} &= A_7 + A^P_4 + A^P_6+ A^P_7 + 3A^P_{10},\qquad A_8^{T+P} = A_8 + A^P_1 - A^P_3+ A^P_8 + 3A^P_{11}, \nonumber\\
  A_9^{T+P} &= A_9 + A^P_2 + A^P_5+ A^P_9 + 3A^P_{12}.
\end{align}
Ultimately, there are eight independent amplitudes contributing to the $\mathcal{B}_{cc}\to \mathcal{B}_{c\overline 3}M$ decays.
In the $SU(3)_F$ limit, amplitudes $A_{7,8,9}^{T+P}$ are negligible in branching fractions, since $|V_{cb}^*V_{ub}| \ll |V_{cd}^*V_{ud}|$ and $|V_{cs}^*V_{us}|$.
Thus, there are five independent amplitudes dominating the branching fractions.
However, amplitudes $A_{7,8,9}^{T+P}$ are not negligible in $CP$ asymmetries due to the large weak phase of the CKM matrix element $V_{ub}$.

The $SU(3)$ irreducible representation amplitude for the $\mathcal{B}_{cc}\to \mathcal{B}_{c6}M$ decay is given by
\begin{align}
  \mathcal{A}^{IR}(\mathcal{B}_{cc}\to \mathcal{B}_{c6}M) &= a^\prime_1(\mathcal{B}_{cc})_iH(15)^i_{jk}M^j_l\mathcal{B}_{c6}^{kl}+ a^\prime_2(\mathcal{B}_{cc})_iH(15)^l_{jk}M^k_l\mathcal{B}_{c6}^{ij}
  +a^\prime_3(\mathcal{B}_{cc})_iH(\overline 6)^i_{jk}M^j_l\mathcal{B}_{c6}^{kl}\nonumber\\&
  +a^\prime_4(\mathcal{B}_{cc})_iH(\overline 6)^l_{jk}M^k_l\mathcal{B}_{c6}^{ij}+ a^\prime_5(\mathcal{B}_{cc})_iH(15)^j_{lk}M^i_j\mathcal{B}_{c6}^{lk}+  a^\prime_6(\mathcal{B}_{cc})_iH(15)^i_{jk}M^l_l\mathcal{B}_{c6}^{jk}\nonumber\\
   &
   + a^\prime_7(\mathcal{B}_{cc})_iH(3)_{j}M^j_k\mathcal{B}_{c6}^{ik}+ a^\prime_8(\mathcal{B}_{cc})_iH(3)_{j}M^i_k\mathcal{B}_{c6}^{jk}
   +a^\prime_9(\mathcal{B}_{cc})_iH(3)_{j}M^k_k\mathcal{B}_{c6}^{ij}
   \nonumber\\&+ a^\prime_{10}(\mathcal{B}_{cc})_iH(3^\prime)_{j}M^j_k\mathcal{B}_{c6}^{ik}+ a^\prime_{11}(\mathcal{B}_{cc})_iH(3^\prime)_{j}M^i_k\mathcal{B}_{c6}^{jk}
   +a^\prime_{12}(\mathcal{B}_{cc})_iH(3^\prime)_{j}M^k_k\mathcal{B}_{c6}^{ij}.
\end{align}
According to Eq.~\eqref{su}, the relations between the $SU(3)$ irreducible representation amplitudes and the topological amplitudes in $\mathcal{B}_{cc}\to \mathcal{B}_{c6}M$ decays are derived as
\begin{align}\label{sol2}
   & a^\prime_1 = A^\prime_1 + A^\prime_6,\qquad a^\prime_2 = A^\prime_4 + A^\prime_5,  \qquad a^\prime_3 = A^\prime_1 + A^\prime_6,  \qquad a^\prime_4 = A^\prime_4 - A^\prime_5, \qquad a^\prime_5 = A^\prime_3 \qquad a^\prime_6 = A^\prime_2, \nonumber\\
   &  a^\prime_7 = \frac{3}{8}A^\prime_1 -\frac{1}{8}A^\prime_4 -\frac{1}{8}A^\prime_6 + A^\prime_7,\quad a^\prime_8 = -\frac{1}{8}A^\prime_1 +  \frac{1}{4}A^\prime_3 +  \frac{3}{8}A^\prime_5 +\frac{3}{8}A^\prime_6  + A^\prime_8,\quad  a^\prime_9 = \frac{1}{4}A^\prime_2 +  \frac{3}{8}A^\prime_4- \frac{1}{8}A^\prime_5 + A^\prime_9,\nonumber\\
    &  a^\prime_{10} = -\frac{1}{8}A^\prime_1 +\frac{3}{8}A^\prime_4 +  \frac{3}{8}A^\prime_6 + A^\prime_{10},\quad a^\prime_{11} = \frac{3}{8}A^\prime_1 +\frac{1}{4}A^\prime_3 -  \frac{1}{8}A^\prime_5 -\frac{1}{8}A^\prime_6  + A^\prime_{11},\nonumber\\ &  a^\prime_{12} = \frac{1}{4}A^\prime_2 -  \frac{1}{8}A^\prime_4+ \frac{3}{8}A^\prime_5 + A^\prime_{12}.
\end{align}
The penguin-induced diagrams appear together with tree induced diagrams $A^{\prime}_{7}$, $A^{\prime}_8$, and $A^{\prime}_9$ with the fixed combinations:
\begin{align}
 A_7^{
 \prime T+P} &= A^\prime_7 + A^{\prime P}_4 + A^{\prime P}_6+ A^{\prime P}_7 + 3A^{\prime P}_{10},\qquad A_8^{T+P} = A^\prime_8 + A^{\prime P}_1 + A^{\prime P}_3+ A^{\prime P}_8 + 3A^{\prime P}_{11}, \nonumber\\
  A_9^{\prime T+P} &= A^\prime_9 + A^{\prime P}_2 + A^{\prime P}_5+ A^{\prime P}_9 + 3A^{\prime P}_{12}.
\end{align}
Thus, there are nine independent amplitudes contributing to the $\mathcal{B}_{cc}\to \mathcal{B}_{c6}M$ decays, and six of them dominate the branching fractions.

\subsection{Decays into a light baryon and a $D$ meson}\label{MB}
In this section, we study the topological diagrams of charmed baryon decays into a light baryon and a charmed meson.
The charmed meson anti-triplet is
\begin{align}
  D = (D^0,D^+,D^+_s).
\end{align}
The light baryons form an $SU(3)$ octet and a decuplet.
According to the Pauli exclusion principle, the combination of flavor and spin wavefunctions must be symmetric under the interchange of any two quarks in the baryon.
The combinations of mixed symmetry wavefunctions, $\phi_S\chi_S$ and $\phi_A\chi_A$, are
both symmetric under the interchange of $q_1\leftrightarrow q_2$.
However, neither combination on its own has a definite symmetry under the interchange of $q_1\leftrightarrow q_3$ or $q_2\leftrightarrow q_3$.
Only their equally weighted linear combination, $\phi_S\chi_S+\phi_A\chi_A$,
is symmetric under the interchange of any two quarks.
Because of the different spin structures of two octets $\phi_S$ and $\phi_A$, the topological diagrams of doubly charmed baryon decays into octet baryons have two independent sets.
For distinction, we designate these two baryon octets as $\mathcal{B}^S_8$ and $\mathcal{B}^A_8$.
The light baryon octet is given as
\begin{align}
  \Sigma^+ & = \frac{1}{\sqrt{6}}(2uus-suu-usu),\qquad   p =\frac{1}{\sqrt{6}}(-2uud+duu+udu), \qquad   \Sigma^- = \frac{1}{\sqrt{6}}(-2dds+sdd+dsd), \nonumber\\
    n &=\frac{1}{\sqrt{6}}(2ddu-udd-dud), \qquad\Xi^- = \frac{1}{\sqrt{6}}(2ssd-dss-sds), \qquad   \Xi^0 =\frac{1}{\sqrt{6}}(-2ssu+uss+sus),\nonumber\\
 \Sigma^0 &=\frac{1}{\sqrt{12}}(sdu+dsu+sud
 +usd-2uds-2dus), \qquad \Lambda^0 = \frac{1}{2}(usd-dsu
 +sud-sdu).
\end{align}
The light baryon octet is given as
\begin{align}
  \Sigma^+ & = \frac{1}{\sqrt{2}}(suu-usu),\qquad   p =\frac{1}{\sqrt{2}}(udu-duu), \qquad \Sigma^- = \frac{1}{\sqrt{2}}(dsd-sdd),  \nonumber\\
  n& =\frac{1}{\sqrt{2}}(udd-dud), \qquad  \Xi^- = \frac{1}{\sqrt{2}}(uss-sus),\qquad   \Xi^0 =\frac{1}{\sqrt{2}}(sus-uss),  \nonumber\\
\Sigma^0 &=\frac{1}{2}(usd-sud+dsu-sdu),   \qquad   \Lambda^0  = \frac{1}{\sqrt{12}}(2dus-2uds+dsu
-sdu+sud-usd).
\end{align}

\begin{figure}
  \centering
  \includegraphics[width=11cm]{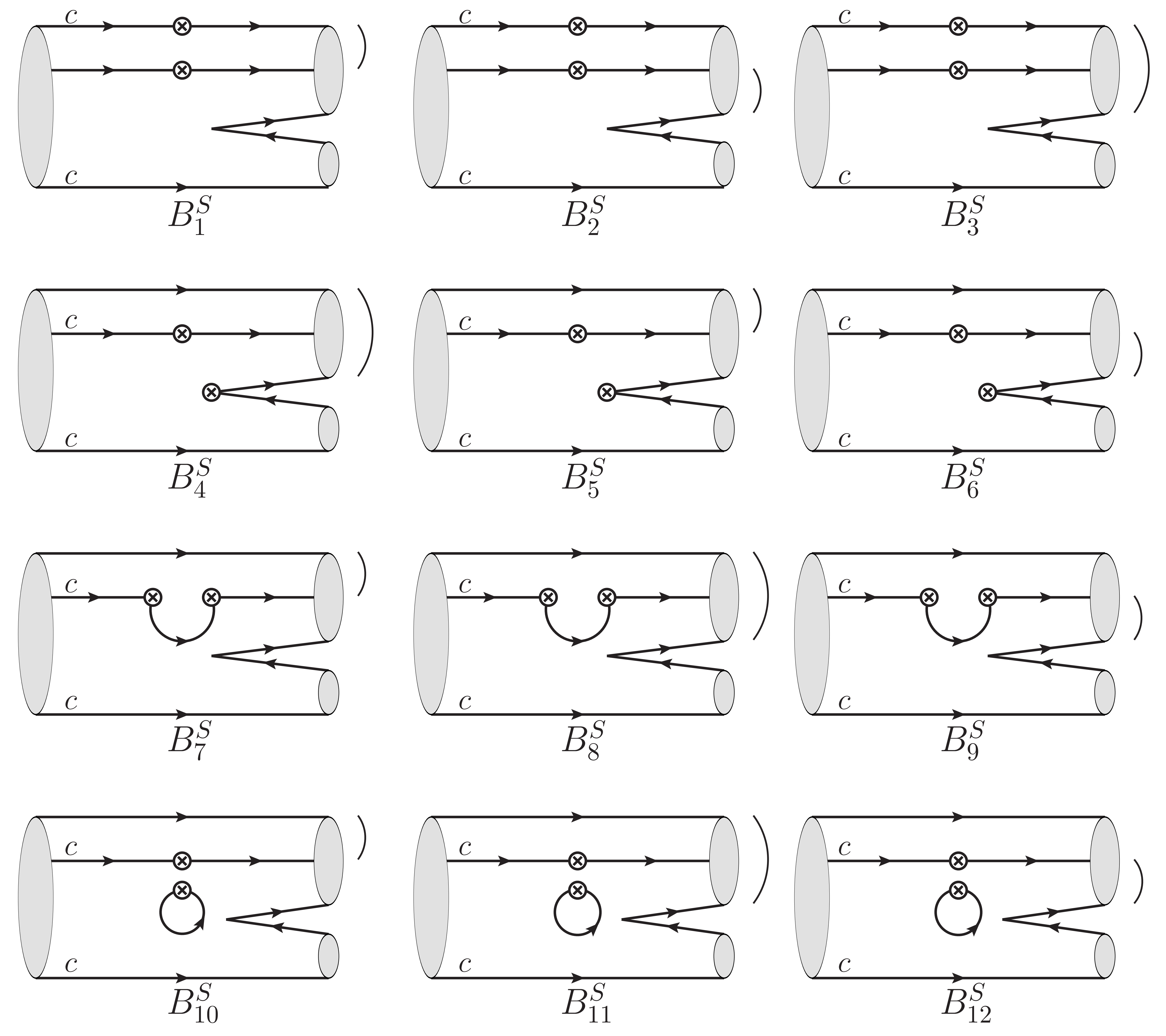}
  \caption{Topological diagrams in doubly charmed baryon decays into a charmed meson and a light octet baryon $\mathcal{B}_8^S$, in which "$)$" indicates that the two light quarks in final-state baryon are symmetric in flavor.}\label{top3}
\end{figure}

The amplitude of $\mathcal{B}_{cc}\to \mathcal{B}^S_8 D$ can be expressed as
\begin{align}\label{a2}
  \mathcal{A}^S(\mathcal{B}_{cc}\to \mathcal{B}^S_8 D) &= B^S_1 (\mathcal{B}_{cc})_i H^i_{jk}D_l (\mathcal{B}_8^S)^{jkl} +B^S_2 (\mathcal{B}_{cc})_i H^i_{jk}D_l (\mathcal{B}_8^S)^{jlk} +  B^S_3 (\mathcal{B}_{cc})_i H^i_{jk}D_l (\mathcal{B}_8^S)^{klj} \nonumber\\
 & + B^S_4 (\mathcal{B}_{cc})_i H^l_{jk}D_l (\mathcal{B}_8^S)^{ijk}+ B^S_5 (\mathcal{B}_{cc})_i H^l_{jk}D_l (\mathcal{B}_8^S)^{ikj} + B^S_6 (\mathcal{B}_{cc})_i H^l_{jk}D_l (\mathcal{B}_8^S)^{jki}\nonumber\\
  & + B^S_7 (\mathcal{B}_{cc})_i H^l_{jl}D_k (\mathcal{B}_8^S)^{ijk} + B^S_8 (\mathcal{B}_{cc})_i H^l_{jl}D_k (\mathcal{B}_8^S)^{ikj}+ B^S_9 (\mathcal{B}_{cc})_i H^l_{jl}D_k (\mathcal{B}_8^S)^{jki}\nonumber\\
  & + B^S_{10} (\mathcal{B}_{cc})_i H^l_{lj}D_k (\mathcal{B}_8^S)^{ijk} + B^S_{11} (\mathcal{B}_{cc})_i H^l_{lj}D_k (\mathcal{B}_8^S)^{ikj}+ B^S_{12} (\mathcal{B}_{cc})_i H^l_{lj}D_k (\mathcal{B}_8^S)^{jki}.
\end{align}
The amplitude of $\mathcal{B}_{cc}\to \mathcal{B}^A_8 D$ can be expressed as
\begin{align}\label{a3}
  \mathcal{A}^{A}(\mathcal{B}_{cc}\to \mathcal{B}^A_8 D) &=  B^A_1 (\mathcal{B}_{cc})_i H^i_{jk}D_l (\mathcal{B}_8^A)^{jkl} +B^A_2 (\mathcal{B}_{cc})_i H^i_{jk}D_l (\mathcal{B}_8^A)^{jlk} +  B^A_3 (\mathcal{B}_{cc})_i H^i_{jk}D_l (\mathcal{B}_8^A)^{klj} \nonumber\\
 & + B^A_4 (\mathcal{B}_{cc})_i H^l_{jk}D_l (\mathcal{B}_8^A)^{ijk}+ B^A_5 (\mathcal{B}_{cc})_i H^l_{jk}D_l (\mathcal{B}_8^A)^{ikj} + B^A_6 (\mathcal{B}_{cc})_i H^l_{jk}D_l (\mathcal{B}_8^A)^{jki}\nonumber\\
  & + B^A_7 (\mathcal{B}_{cc})_i H^l_{jl}D_k (\mathcal{B}_8^A)^{ijk} + B^A_8 (\mathcal{B}_{cc})_i H^l_{jl}D_k (\mathcal{B}_8^A)^{ikj}+ B^A_9 (\mathcal{B}_{cc})_i H^l_{jl}D_k (\mathcal{B}_8^A)^{jki}\nonumber\\
  & + B^A_{10} (\mathcal{B}_{cc})_i H^l_{lj}D_k (\mathcal{B}_8^A)^{ijk} + B^A_{11} (\mathcal{B}_{cc})_i H^l_{lj}D_k (\mathcal{B}_8^A)^{ikj}+ B^A_{12} (\mathcal{B}_{cc})_i H^l_{lj}D_k (\mathcal{B}_8^A)^{jki}.
\end{align}
The total number of amplitudes contributing to the $\mathcal{B}_{cc}\to \mathcal{B}^S_8 D$ and $\mathcal{B}_{cc}\to \mathcal{B}^A_8 D$ modes is $N_S+N_A=A^4_4=24$.
The total amplitude of $\mathcal{B}_{cc}\to \mathcal{B}_8 D$ decay is given by
\begin{align}
  \mathcal{A}(\mathcal{B}_{cc}\to \mathcal{B}_8 D) = \mathcal{A}^S(\mathcal{B}_{cc}\to \mathcal{B}^S_8 D) + \mathcal{A}^A(\mathcal{B}_{cc}\to \mathcal{B}^S_8 D).
\end{align}
The topological diagrams contributing to the $\mathcal{B}_{cc}\to \mathcal{B}^S_8 D$ decays are shown in Fig.~\ref{top3}.
The topological diagrams contributing to the $\mathcal{B}_{cc}\to \mathcal{B}^A_8 D$ decays can be obtained by replacing the symmetric quarks with the antisymmetric ones.
The topological amplitudes of $\mathcal{B}_{cc}\to \mathcal{B}^S_8 D$ and $\mathcal{B}_{cc}\to \mathcal{B}^A_8 D$ decays are listed in Tables~\ref{amp4} and \ref{amp5}, respectively.

\begin{table*}
\caption{Topological amplitudes of $\mathcal{B}_{cc}\to \mathcal{B}_{8}^SD$ decays.}\label{amp4}
 \small
\begin{tabular}{|c|c|c|c|}
\hline\hline
 Channel & Amplitude & Channel &Amplitude\\\hline
 $\Xi_{cc}^{++}\to \Sigma^{+}D^+$ & $\frac{1}{\sqrt{6}}\lambda_1(2B^S_4-B^S_5-B^S_6)$ &$\Xi_{cc}^{++}\to p D^+_s$ & $\frac{1}{\sqrt{6}}\lambda_2(-2B^S_4+B^S_5+B^S_6)$ \\\hline
 $\Xi_{cc}^{+}\to \Lambda^{0}D^+$ & $\frac{1}{2}\lambda_1(-B^S_1+B^S_3+B^S_5-B^S_6)$ &$\Xi_{cc}^{+}\to n D^+_s$ & $\frac{1}{\sqrt{6}}\lambda_2(-B^S_4+2B^S_5-B^S_6)$ \\\hline
 $\Xi_{cc}^{+}\to \Sigma^{+}D^0$ & $\frac{1}{\sqrt{6}}\lambda_1(-B^S_1+2B^S_2-B^S_3)$ &$\Omega_{cc}^{+}\to \Lambda^{0} D^+_s$ & $\frac{1}{2}\lambda_2(-B^S_2+B^S_3-B^S_4+B^S_5)$ \\\hline
 $\Xi_{cc}^{+}\to \Sigma^{0}D^+$ & ~$\frac{1}{2\sqrt{3}}\lambda_1(B^S_1-2B^S_2+B^S_3-2B^S_4+B^S_5+B^S_6)$~ &$\Omega_{cc}^{+}\to \Sigma^{0} D^+_s$ & $\frac{1}{2\sqrt{3}}\lambda_2(-2B^S_1+B^S_2+B^S_3+B^S_4+B^S_5-2B^S_6)$ \\\hline
 $\Xi_{cc}^{+}\to \Xi^{0}D^+_s$ &$\frac{1}{\sqrt{6}}\lambda_1(B^S_1+B^S_2-2B^S_3)$  &$\Omega_{cc}^{+}\to p D^0$ &  $\frac{1}{\sqrt{6}}\lambda_2(B^S_1-2B^S_2+B^S_3)$ \\\hline
 $\Omega_{cc}^{+}\to \Xi^{0}D^+$ & $\frac{1}{\sqrt{6}}\lambda_1(B^S_4-2B^S_5+B^S_6)$ &$\Omega_{cc}^{+}\to n D^+$ & $\frac{1}{\sqrt{6}}\lambda_2(-B^S_1-B^S_2+2B^S_3)$ \\\toprule[1.2pt]
$\Xi_{cc}^{++}\to \Sigma^{+}D^+_s$ &  \multicolumn{3}{c|}{ $\frac{1}{\sqrt{6}}\lambda_d(2B^S_7-B^S_8-B^S_9)+\frac{1}{\sqrt{6}}\lambda_s(2B^S_4-B^S_5-B^S_6+2B^S_7-B^S_8-B^S_9)$ } \\ & \multicolumn{3}{c|}{ $-\frac{1}{\sqrt{6}}\lambda_b(2B^{SP}_1-B^{SP}_2-B^{SP}_3-B^{SP}_4+2B^{SP}_5-B^{SP}_6+2B^{SP}_7-B^{SP}_8-B^{SP}_9+6B^{SP}_{10}-3B^{SP}_{11}-3B^{SP}_{12})$ }\\\hline
$\Xi_{cc}^{++}\to pD^+$ &  \multicolumn{3}{c|}{ $\frac{1}{\sqrt{6}}\lambda_d(-2B^S_4+B^S_5+B^S_6-2B^S_7+B^S_8+B^S_9)+\frac{1}{\sqrt{6}}\lambda_s(-2B^S_7+B^S_8+B^S_9)$ } \\ & \multicolumn{3}{c|}{ $-\frac{1}{\sqrt{6}}\lambda_b(-2B^{SP}_1+B^{SP}_2+B^{SP}_3+B^{SP}_4-2B^{SP}_5+B^{SP}_6-2B^{SP}_7+B^{SP}_8+B^{SP}_9-6B^{SP}_{10}+3B^{SP}_{11}+3B^{SP}_{12})$ }\\\hline
$\Xi_{cc}^{+}\to \Lambda^{0}D^+_s$ &  \multicolumn{3}{c|}{ $\frac{1}{2}\lambda_d(-B^S_2+B^S_3+B^S_8-B^S_9)+\frac{1}{2}\lambda_s(B^S_5-B^S_6+B^S_8-B^S_9)$ } \\ & \multicolumn{3}{c|}{ $-\frac{1}{2}\lambda_b(B^{SP}_2-B^{SP}_3+B^{SP}_4-B^{SP}_6+B^{SP}_8-B^{SP}_9+3B^{SP}_{11}-3B^{SP}_{12})$ }\\\hline
$\Xi_{cc}^{+}\to \Sigma^{0}D^+_s$ &  \multicolumn{3}{c|}{ $\frac{1}{2\sqrt{3}}\lambda_d(-2B^S_1+B^S_2+B^S_3-2B^S_7+B^S_8+B^S_9)+\frac{1}{2\sqrt{3}}\lambda_s(-2B^S_4+B^S_5+B^S_6-2B^S_7+B^S_8+B^S_9)$ } \\ & \multicolumn{3}{c|}{ $-\frac{1}{2\sqrt{3}}\lambda_b(-2B^{SP}_1+B^{SP}_2+B^{SP}_3+B^{SP}_4-2B^{SP}_5+B^{SP}_6-2B^{SP}_7+B^{SP}_8+B^{SP}_9-6B^{SP}_{10}+3B^{SP}_{11}+3B^{SP}_{12})$ }\\\hline
$\Xi_{cc}^{+}\to pD^0$ &  \multicolumn{3}{c|}{ $\frac{1}{\sqrt{6}}\lambda_d(B^S_1-2B^S_2+B^S_3+B^S_7+B^S_8-2B^S_9)+\frac{1}{\sqrt{6}}\lambda_s(B^S_7+B^S_8-2B^S_9)$ } \\ & \multicolumn{3}{c|}{ $-\frac{1}{\sqrt{6}}\lambda_b(B^{SP}_1+B^{SP}_2-2B^{SP}_3+B^{SP}_4+B^{SP}_5-2B^{SP}_6+B^{SP}_7+B^{SP}_8-2B^{SP}_9+3B^{SP}_{10}+3B^{SP}_{11}-6B^{SP}_{12})$ }\\\hline
$\Xi_{cc}^{+}\to nD^+$ &  \multicolumn{3}{c|}{ $\frac{1}{\sqrt{6}}\lambda_d(-B^S_1-B^S_2+2B^S_3-B^S_4+2B^S_5-B^S_6-B^S_7+2B^S_8-B^S_9)+\frac{1}{\sqrt{6}}\lambda_s(-B^S_7+2B^S_8-B^S_9)$ } \\ & \multicolumn{3}{c|}{ $-\frac{1}{\sqrt{6}}\lambda_b(-B^{SP}_1+2B^{SP}_2-B^{SP}_3+2B^{SP}_4-B^{SP}_5-B^{SP}_6-B^{SP}_7+2B^{SP}_8-B^{SP}_9-3B^{SP}_{10}+6B^{SP}_{11}-3B^{SP}_{12})$ }\\\hline
$\Omega_{cc}^{+}\to \Lambda^{0}D^+$ &  \multicolumn{3}{c|}{ $\frac{1}{2}\lambda_d(-B^S_4+B^S_5-B^S_7+B^S_8)+\frac{1}{2}\lambda_s(-B^S_1+B^S_3-B^S_7+B^S_8)$ } \\ & \multicolumn{3}{c|}{ $-\frac{1}{2}\lambda_b(-B^{SP}_1+B^{SP}_2+B^{SP}_4-B^{SP}_5-B^{SP}_7+B^{SP}_8-3B^{SP}_{10}+3B^{SP}_{11})$ }\\\hline
$\Omega_{cc}^{+}\to \Sigma^{+}D^0$ &  \multicolumn{3}{c|}{ $\frac{1}{\sqrt{6}}\lambda_d(-B^S_7-B^S_8+2B^S_9)+\frac{1}{\sqrt{6}}\lambda_s(-B^S_1+2B^S_2-B^S_3-B^S_7-B^S_8+2B^S_9)$ } \\ & \multicolumn{3}{c|}{ $-\frac{1}{\sqrt{6}}\lambda_b(-B^{SP}_1-B^{SP}_2+2B^{SP}_3-B^{SP}_4-B^{SP}_5+2B^{SP}_6-B^{SP}_7-B^{SP}_8+2B^{SP}_9-3B^{SP}_{10}-3B^{SP}_{11}+6B^{SP}_{12})$ }\\\hline
$\Omega_{cc}^{+}\to \Sigma^{0}D^+$ &  \multicolumn{3}{c|}{ $\frac{1}{2\sqrt{3}}\lambda_d(B^S_4+B^S_5-2B^S_6+B^S_7+B^S_8-2B^S_9)+\frac{1}{2\sqrt{3}}\lambda_s(B^S_1-2B^S_2+B^S_3+B^S_7+B^S_8-2B^S_9)$ } \\ & \multicolumn{3}{c|}{ $-\frac{1}{2\sqrt{3}}\lambda_b(B^{SP}_1+B^{SP}_2-2B^{SP}_3+B^{SP}_4+B^{SP}_5-2B^{SP}_6+B^{SP}_7+B^{SP}_8-2B^{SP}_9+3B^{SP}_{10}+3B^{SP}_{11}-6B^{SP}_{12})$ }\\\hline
$\Omega_{cc}^{+}\to \Xi^{0}D^+_s$ &  \multicolumn{3}{c|}{ $\frac{1}{\sqrt{6}}\lambda_d(B^S_7-2B^S_8+B^S_9)+\frac{1}{\sqrt{6}}\lambda_s(B^S_1+B^S_2-2B^S_3+B^S_4-2B^S_5+B^S_6+B^S_7-2B^S_8+B^S_9)$ } \\ & \multicolumn{3}{c|}{ $-\frac{1}{\sqrt{6}}\lambda_b(B^{SP}_1-2B^{SP}_2+B^{SP}_3-2B^{SP}_4+B^{SP}_5+B^{SP}_6+B^{SP}_7-2B^{SP}_8+B^{SP}_9+3B^{SP}_{10}-6B^{SP}_{11}+3B^{SP}_{12})$ }\\\hline
  \hline
\end{tabular}
\end{table*}

\begin{table*}
\caption{Topological amplitudes of $\mathcal{B}_{cc}\to \mathcal{B}_{8}^AD$ decays.}\label{amp5}
 \small
\begin{tabular}{|c|c|c|c|}
\hline\hline
 Channel & Amplitude & Channel &Amplitude\\\hline
 $\Xi_{cc}^{++}\to \Sigma^{+}D^+$ & $-\frac{1}{\sqrt{2}}\lambda_1(B^A_5+B^A_6)$ &$\Xi_{cc}^{++}\to p D^+_s$ & $\frac{1}{\sqrt{2}}\lambda_2(B^A_5+B^A_6)$ \\\hline
 $\Xi_{cc}^{+}\to \Lambda^{0}D^+$ & $-\frac{1}{2\sqrt{3}}\lambda_1(B^A_1+2B^A_2+B^A_3-2B^A_4-B^A_5+B^A_6)$ &$\Xi_{cc}^{+}\to n D^+_s$ & $-\frac{1}{\sqrt{2}}\lambda_2(B^A_4-B^A_6)$ \\\hline
 $\Xi_{cc}^{+}\to \Sigma^{+}D^0$ & $\frac{1}{\sqrt{2}}\lambda_1(-B^A_1+B^A_3)$ &$\Omega_{cc}^{+}\to \Lambda^{0} D^+_s$ &$-\frac{1}{2\sqrt{3}}\lambda_2(2B^A_1+B^A_2-B^A_3-B^A_4+B^A_5+2B^A_6)$ \\\hline
 $\Xi_{cc}^{+}\to \Sigma^{0}D^+$ & $\frac{1}{2}\lambda_1(B^A_1-B^A_3+B^A_5+B^A_6)$ &$\Omega_{cc}^{+}\to \Sigma^{0} D^+_s$ &$\frac{1}{2}\lambda_2(B^A_2+B^A_3-B^A_4-B^A_5)$  \\\hline
 $\Xi_{cc}^{+}\to \Xi^{0}D^+_s$ & $-\frac{1}{\sqrt{2}}\lambda_1(B^A_1+B^A_2)$ &$\Omega_{cc}^{+}\to p D^0$ &$\frac{1}{\sqrt{2}}\lambda_2(B^A_1-B^A_3)$ \\\hline
 $\Omega_{cc}^{+}\to \Xi^{0}D^+$ & $\frac{1}{\sqrt{2}}\lambda_1(B^A_4-B^A_6)$ &$\Omega_{cc}^{+}\to n D^+$ & $\frac{1}{\sqrt{2}}\lambda_2(B^A_1+B^A_2)$ \\\toprule[1.2pt]
$\Xi_{cc}^{++}\to \Sigma^{+}D^+_s$ &  \multicolumn{3}{c|}{$-\frac{1}{\sqrt{2}}\lambda_d(B^A_8+B^A_9)-\frac{1}{\sqrt{2}}\lambda_s(B^A_5+B^A_6+B^A_8+B^A_9)$ }\\ & \multicolumn{3}{c|}{$+\frac{1}{\sqrt{2}}\lambda_b(B^{AP}_2+B^{AP}_3+B^{AP}_4-B^{AP}_6+B^{AP}_8+B^{AP}_9+3B^{AP}_{11}+3B^{AP}_{12})$} \\\hline
$\Xi_{cc}^{++}\to pD^+$ & \multicolumn{3}{c|}{$\frac{1}{\sqrt{2}}\lambda_d(B^A_5+B^A_6+B^A_8+B^A_9)+\frac{1}{\sqrt{2}}\lambda_s(B^A_8+B^A_9)$ }\\ & \multicolumn{3}{c|}{$-\frac{1}{\sqrt{2}}\lambda_b(B^{AP}_2+B^{AP}_3+B^{AP}_4-B^{AP}_6+B^{AP}_8+B^{AP}_9+3B^{AP}_{11}+3B^{AP}_{12})$}  \\\hline
$\Xi_{cc}^{+}\to \Lambda^{0}D^+_s$ & \multicolumn{3}{c|}{$-\frac{1}{2\sqrt{3}}\lambda_d(2B^A_1+B^A_2-B^A_3-2B^A_7-B^A_8+
B^A_9)-\frac{1}{2\sqrt{3}}\lambda_s(-2B^A_4-B^A_5+B^A_6-2B^A_7-B^A_8+B^A_9)$ }\\ & \multicolumn{3}{c|}{$+\frac{1}{2\sqrt{3}}\lambda_b(-2B^{AP}_1-B^{AP}_2+B^{AP}_3-B^{AP}_4-2B^{AP}_5-B^{AP}_6-2B^{AP}_7-B^{AP}_8+B^{AP}_9-6B^{AP}_{10}-3B^{AP}_{11}+3B^{AP}_{12})$}  \\\hline
$\Xi_{cc}^{+}\to \Sigma^{0}D^+_s$ & \multicolumn{3}{c|}{$\frac{1}{2}\lambda_d(B^A_2+B^A_3+B^A_8+B^A_9)+\frac{1}{2}\lambda_s(B^A_6+B^A_8+B^A_9)$ }\\ & \multicolumn{3}{c|}{$-\frac{1}{2}\lambda_b(B^{AP}_2+B^{AP}_3+B^{AP}_4-B^{AP}_6+B^{AP}_8+B^{AP}_9+3B^{AP}_{11}+3B^{AP}_{12})$}  \\\hline
$\Xi_{cc}^{+}\to pD^0$ &  \multicolumn{3}{c|}{$-\frac{1}{\sqrt{2}}\lambda_d(-B^A_1+B^A_3+B^A_7+B^A_8)-\frac{1}{\sqrt{2}}\lambda_s(B^A_7+B^A_8)$ }\\ & \multicolumn{3}{c|}{$+\frac{1}{\sqrt{2}}\lambda_b(B^{AP}_1+B^{AP}_2+B^{AP}_4+B^{AP}_5+B^{AP}_7+B^{AP}_8+3B^{AP}_{10}+3B^{AP}_{12})$}  \\\hline
$\Xi_{cc}^{+}\to nD^+$ &  \multicolumn{3}{c|}{$\frac{1}{\sqrt{2}}\lambda_d(B^A_1+B^A_2-B^A_4+B^A_6-B^A_7+B^A_9)+\frac{1}{\sqrt{2}}\lambda_s(-B^A_7+B^A_9)$ }\\ & \multicolumn{3}{c|}{$+\frac{1}{\sqrt{2}}\lambda_b(B^{AP}_1-B^{AP}_3+B^{AP}_5+B^{AP}_6+B^{AP}_7-B^{AP}_9+3B^{AP}_{10}-3B^{AP}_{12})$}  \\\hline
$\Omega_{cc}^{+}\to \Lambda^{0}D^+$ &  \multicolumn{3}{c|}{$-\frac{1}{2\sqrt{3}}\lambda_d(-B^A_4+B^A_5+2B^A_6-B^A_7+B^A_8+
2B^A_9)-\frac{1}{2\sqrt{3}}\lambda_s(B^A_1+2B^A_2+B^A_3-B^A_7+B^A_8+2B^A_9)$ }\\ & \multicolumn{3}{c|}{$+\frac{1}{2\sqrt{3}}\lambda_b(-B^{AP}_1+B^{AP}_2+2B^{AP}_3+B^{AP}_4-B^{AP}_5-2B^{AP}_6-B^{AP}_7+B^{AP}_8+2B^{AP}_9-3B^{AP}_{10}+3B^{AP}_{11}+6B^{AP}_{12})$}  \\\hline
$\Omega_{cc}^{+}\to \Sigma^{+}D^0$ &  \multicolumn{3}{c|}{$\frac{1}{\sqrt{2}}\lambda_d(B^A_7+B^A_8)+\frac{1}{\sqrt{2}}\lambda_s(-B^A_1+B^A_3+B^A_7+B^A_8)$ }\\ & \multicolumn{3}{c|}{$-\frac{1}{\sqrt{2}}\lambda_b(B^{AP}_1+B^{AP}_2+B^{AP}_4+B^{AP}_5+B^{AP}_7+B^{AP}_8+3B^{AP}_{10}+3B^{AP}_{11})$}  \\\hline
$\Omega_{cc}^{+}\to \Sigma^{0}D^+$ &  \multicolumn{3}{c|}{$-\frac{1}{2}\lambda_d(B^A_4+B^A_5+B^A_7+B^A_8)-\frac{1}{2}\lambda_s(-B^A_1+B^A_3+B^A_7+B^A_8)$ }\\ & \multicolumn{3}{c|}{$+\frac{1}{2}\lambda_b(B^{AP}_1+B^{AP}_2+B^{AP}_4+B^{AP}_5+B^{AP}_7+B^{AP}_8+3B^{AP}_{10}+3B^{AP}_{11})$}  \\\hline
$\Omega_{cc}^{+}\to \Xi^{0}D^+_s$ &  \multicolumn{3}{c|}{$\frac{1}{\sqrt{2}}\lambda_d(B^A_7-B^A_9)-\frac{1}{\sqrt{2}}\lambda_s(B^A_1+B^A_2-B^A_4+B^A_6-B^A_7+B^A_9)$ }\\ & \multicolumn{3}{c|}{$-\frac{1}{\sqrt{2}}\lambda_b(B^{AP}_1-B^{AP}_3+B^{AP}_5+B^{AP}_6+B^{AP}_7-B^{AP}_9+3B^{AP}_{10}-3B^{AP}_{12})$}  \\\hline
  \hline
\end{tabular}
\end{table*}

\begin{table*}
\caption{Decay amplitudes of $\mathcal{B}_{cc}\to \mathcal{B}_{8}D$ modes constructed by $(1,1)$-rank baryon octet.}\label{ampb}
 \small
\begin{tabular}{|c|c|c|c|}
\hline\hline
 Channel & \qquad\qquad\quad Amplitude \qquad\qquad\qquad & Channel &Amplitude\\\hline
 $\Xi_{cc}^{++}\to \Sigma^{+}D^+$ & $-\lambda_1(b_4+b_5)$ &$\Xi_{cc}^{++}\to p D^+_s$ & $\lambda_2(b_4+b_5)$ \\\hline
 $\Xi_{cc}^{+}\to \Lambda^{0}D^+$ & $\frac{1}{\sqrt{6}}\lambda_1(2b_1+b_2-b_5)$ &$\Xi_{cc}^{+}\to n D^+_s$ & $\lambda_2(-b_3+b_5)$ \\\hline
 $\Xi_{cc}^{+}\to \Sigma^{+}D^0$ & $\lambda_1(-b_2+b_4)$ &$\Omega_{cc}^{+}\to \Lambda^{0} D^+_s$ &$\frac{1}{\sqrt{6}}\lambda_2(b_1-b_2-2b_5)$ \\\hline
 $\Xi_{cc}^{+}\to \Sigma^{0}D^+$ & $\frac{1}{\sqrt{2}}\lambda_1(b_2+b_5)$ &$\Omega_{cc}^{+}\to \Sigma^{0} D^+_s$ &$-\frac{1}{\sqrt{2}}\lambda_2(b_1+b_2)$  \\\hline
 $\Xi_{cc}^{+}\to \Xi^{0}D^+_s$ & $\lambda_1(b_1-b_3)$ &$\Omega_{cc}^{+}\to p D^0$ &$\lambda_2(b_2-b_4)$ \\\hline
 $\Omega_{cc}^{+}\to \Xi^{0}D^+$ & $\lambda_1(b_3-b_5)$ &$\Omega_{cc}^{+}\to n D^+$ & $\lambda_2(-b_1+b_3)$ \\\toprule[1.2pt]
$\Xi_{cc}^{++}\to \Sigma^{+}D^+_s$ &  \multicolumn{3}{c|}{$\lambda_d(b_4+b_5+b_6+b_7)+\lambda_s(b_6+b_7)-\lambda_b
(b^P_1+b^P_2+b_3-b^P_5+b^P_6+b^P_7+3 b^P_8+3 b^P_9)$ } \\\hline
$\Xi_{cc}^{++}\to pD^+$ & \multicolumn{3}{c|}{$-\lambda_d(b_6+b_7)
-\lambda_s(b_4+b_5+b_6+b_7)+\lambda_b(b^P_1+b^P_2+b^P_3-b^P_5+b^P_6+b^P_7+3 b^P_8+3 b^P_9)$ }  \\\hline
$\Xi_{cc}^{+}\to \Lambda^{0}D^+_s$ & \multicolumn{3}{c|}{$\frac{1}{\sqrt{6}}
\lambda_d(b^P_1-b^P_2-b^P_3+b^P_5-b^P_6+b^P_7)+\frac{1}{\sqrt{6}}
\lambda_s(2 b_3-b_6+b_7)$}\\ & \multicolumn{3}{c|}{$+\frac{1}{\sqrt{6}}
\lambda_b(b^P_1-b^P_2+b^P_3+b^P_5+b^P_6-b^P_7+3 b^P_8-3 b^P_9)$ } \\\hline
$\Xi_{cc}^{+}\to \Sigma^{0}D^+_s$ & \multicolumn{3}{c|}{$-\frac{1}{\sqrt{2}}\lambda_d
(b_1+b_2-b_3+b_5+b_6+b_7)
-\frac{1}{\sqrt{2}}\lambda_s(b_6+b_7)$}\\& \multicolumn{3}{c|}{$-\frac{1}{\sqrt{2}}\lambda_b
(b^P_1+b^P_2+b^P_3-b^P_5+b^P_6+b^P_7+3 b^P_8+3 b^P_9)$ } \\\hline
$\Xi_{cc}^{+}\to pD^0$ &  \multicolumn{3}{c|}{$\lambda_d(b_2+b_6)+\lambda_s(b_4+b_6)
-\lambda_b(b^P_1+b^P_3+b^P_6+3 b^P_8)$ } \\\hline
$\Xi_{cc}^{+}\to nD^+$ &  \multicolumn{3}{c|}{$-\lambda_d(b_1+b_7)-\lambda_s(b_5+b_7)
+\lambda_b(b^P_2-b^P_5+b^P_7+3 b^P_9)$ } \\\hline
$\Omega_{cc}^{+}\to \Lambda^{0}D^+$ &  \multicolumn{3}{c|}{$\frac{1}{\sqrt{6}}\lambda_d
(b_3+b_6+2b_7)+\frac{1}{\sqrt{6}}\lambda_s
(2 b^P_1+b^P_2-2 b^P_3+2 b^P_5+b^P_6+2 b^P_7)$ }\\ & \multicolumn{3}{c|}{$-\frac{1}{\sqrt{6}}\lambda_b
(b^P_1+2 b^P_2+b^P_3-2 b^P_5+b^P_6+2 b^P_7+3 b^P_8+6 b^P_9)$}  \\\hline
$\Omega_{cc}^{+}\to \Sigma^{+}D^0$ &  \multicolumn{3}{c|}{$-\lambda_d(b_4+b_6)-\lambda_s(b_2+b_6)
+\lambda_b(b^P_1+b^P_3+b^P_6+3 b^P_8)$ }  \\\hline
$\Omega_{cc}^{+}\to \Sigma^{0}D^+$ &  \multicolumn{3}{c|}{$\frac{1}{\sqrt{2}}\lambda_d(-b_3+b_6)
+\frac{1}{\sqrt{2}}\lambda_s(b_2+b_6)
-\frac{1}{\sqrt{2}}\lambda_b(b^P_1+b^P_3+b^P_6+3 b^P_8)$ }  \\\hline
$\Omega_{cc}^{+}\to \Xi^{0}D^+_s$ &  \multicolumn{3}{c|}{$\lambda_d(b_5+b_7)
+\lambda_s(b_1+b_7)-\lambda_b(b^P_2-b^P_5+b^P_7+3 b^P_9)$ } \\\hline\hline
\end{tabular}
\end{table*}

\begin{table*}
\caption{Topological amplitudes of $\mathcal{B}_{cc}\to \mathcal{B}_{10}D$ decays.}\label{amp3}
 \small
\begin{tabular}{|c|c|c|c|}
\hline\hline
 Channel & Amplitude & Channel &Amplitude\\\hline
 $\Xi_{cc}^{++}\to \Sigma^{*+}D^+$ & $\frac{1}{\sqrt{3}}\lambda_1C_2$ &$\Xi_{cc}^{++}\to \Delta^{+} D^+_s$ & $\frac{1}{\sqrt{3}}\lambda_2C_2$ \\\hline
 $\Xi_{cc}^{+}\to \Sigma^{*+}D^0$ & $\frac{1}{\sqrt{3}}\lambda_1C_1$ &$\Xi_{cc}^{+}\to \Delta^{0} D^+_s$ & $\frac{1}{\sqrt{3}}\lambda_2C_2$ \\\hline
 $\Xi_{cc}^{+}\to \Sigma^{*0}D^+$ & ~~~~~~$\frac{1}{\sqrt{6}}\lambda_1(C_1+C_2)$~~~~~ &$\Omega_{cc}^{+}\to \Delta^{+} D^0$ & $\frac{1}{\sqrt{3}}\lambda_2C_1$ \\\hline
 $\Xi_{cc}^{+}\to \Xi^{*0}D^+_s$ & $\frac{1}{\sqrt{3}}\lambda_1C_1$ &$\Omega_{cc}^{+}\to \Delta^{0} D^+$ & $\frac{1}{\sqrt{3}}\lambda_2C_1$ \\\hline
 $\Omega_{cc}^{+}\to \Xi^{*0}D^+$ &$\frac{1}{\sqrt{3}}\lambda_1C_2$  &$\Omega_{cc}^{+}\to \Sigma^{*0} D^+_s$ & $\frac{1}{\sqrt{6}}\lambda_2(C_1+C_2)$ \\\toprule[1.2pt]
$\Xi_{cc}^{++}\to \Delta^{+}D^+$ &  \multicolumn{3}{c|}{$\frac{1}{\sqrt{3}}\lambda_d(C_2+C_3)+\frac{1}{\sqrt{3}}\lambda_sC_3-\frac{1}{\sqrt{3}}\lambda_b(C^P_1+C^P_2+C^P_3+3C^P_4)$ } \\\hline
$\Xi_{cc}^{++}\to \Sigma^{*+}D^+_s$ &  \multicolumn{3}{c|}{$\frac{1}{\sqrt{3}}\lambda_dC_3+\frac{1}{\sqrt{3}}\lambda_s(C_2+C_3)-\frac{1}{\sqrt{3}}\lambda_b(C^P_1+C^P_2+C^P_3+3C^P_4)$ } \\\hline
$\Xi_{cc}^{+}\to \Delta^{+}D^0$ &  \multicolumn{3}{c|}{ $\frac{1}{\sqrt{3}}\lambda_d(C_1+C_3)+\frac{1}{\sqrt{3}}\lambda_sC_3-\frac{1}{\sqrt{3}}\lambda_b(C^P_1+C^P_2+C^P_3+3C^P_4)$} \\\hline
$\Xi_{cc}^{+}\to \Delta^{0}D^+$ &  \multicolumn{3}{c|}{ $\frac{1}{\sqrt{3}}\lambda_d(C_1+C_2+C_3)+\frac{1}{\sqrt{3}}\lambda_sC_3-\frac{1}{\sqrt{3}}\lambda_b(C^P_1+C^P_2+C^P_3+3C^P_4)$} \\\hline
$\Xi_{cc}^{+}\to \Sigma^{*0}D^+_s$ &  \multicolumn{3}{c|}{ $\frac{1}{\sqrt{6}}\lambda_d(C_1+C_3)+\frac{1}{\sqrt{6}}\lambda_s(C_2+C_3)-\frac{1}{\sqrt{6}}\lambda_b(C^P_1+C^P_2+C^P_3+3C^P_4)$} \\\hline
$\Omega_{cc}^{+}\to \Sigma^{*+}D^0$ &  \multicolumn{3}{c|}{$\frac{1}{\sqrt{3}}\lambda_dC_3+\frac{1}{\sqrt{3}}\lambda_s(C_1+C_3)-\frac{1}{\sqrt{3}}\lambda_b(C^P_1+C^P_2+C^P_3+3C^P_4)$ } \\\hline
$\Omega_{cc}^{+}\to \Sigma^{*0}D^+$ &  \multicolumn{3}{c|}{$\frac{1}{\sqrt{6}}\lambda_d(C_2+C_3)+\frac{1}{\sqrt{6}}\lambda_s(C_1+C_3)-\frac{1}{\sqrt{6}}\lambda_b(C^P_1+C^P_2+C^P_3+3C^P_4)$ } \\\hline
$\Omega_{cc}^{+}\to \Xi^{*0}D^+_s$ &  \multicolumn{3}{c|}{$\frac{1}{\sqrt{3}}\lambda_dC_3+\frac{1}{\sqrt{3}}\lambda_s(C_1+C_2+C_3)-\frac{1}{\sqrt{3}}\lambda_b(C^P_1+C^P_2+C^P_3+3C^P_4)$ } \\\hline
  \hline
\end{tabular}
\end{table*}

The octet baryon can also be written as a tensor with one covariant index and one contravariant index, $(\mathcal{B}_8)^i_j$, where $i\neq j$:
\begin{eqnarray}
 \mathcal{B}_8=  \left( \begin{array}{ccc}
   \frac{1}{\sqrt 2} \Sigma^0+  \frac{1}{\sqrt 6} \Lambda^0    & \Sigma^+  & p \\
    \Sigma^- &   - \frac{1}{\sqrt 2} \Sigma^0+ \frac{1}{\sqrt 6} \Lambda^0   & n \\
    \Xi^- & \Xi^0 & -\sqrt{2/3}\Lambda^0 \\
  \end{array}\right).
\end{eqnarray}
The third-rank tensors $(\mathcal{B}^S_8)^{ijk}$ and $(\mathcal{B}^A_8)^{ijk}$ can be expressed in terms of the ($1,1$)-rank tensors as \cite{Wang:2025bdl,Wang:2024ztg}
\begin{eqnarray}\label{sy1}
(\mathcal{B}^S_8)^{ijk} = \frac{1}{\sqrt{6}}\left[\epsilon^{kil}(\mathcal{B}_8)^j_l
+\epsilon^{kjl}(\mathcal{B}_8)^i_l\right],\qquad\quad  (\mathcal{B}^A_8)^{ijk}= \frac{1}{\sqrt{2}}\epsilon^{ijl}(\mathcal{B}_8)^k_l.
\end{eqnarray}
The indices $i$ and $j$ are symmetric under $i \leftrightarrow j$ in $(\mathcal{B}^S_8)^{ijk}$ and antisymmetric in $(\mathcal{B}^A_8)^{ijk}$.
To study the linear relations between decay amplitudes of $\mathcal{B}_{cc}\to \mathcal{B}_{8}^SD$ and $\mathcal{B}_{cc}\to \mathcal{B}_{8}^AD$ modes, we construct the decay amplitude of $\mathcal{B}_{cc}\to \mathcal{B}_{8}D$ with the ($1,1$)-rank octet tensors as
\begin{align}\label{b1}
  \mathcal{A}(\mathcal{B}_{cc}\to \mathcal{B}_8 D) &=  b_1 (\mathcal{B}_{cc})_i H^i_{jk}D^{jl} (\mathcal{B}_8)^{k}_l +b_2 (\mathcal{B}_{cc})_i H^i_{jk}D^{kl} (\mathcal{B}_8)^{j}_l + b_3 (\mathcal{B}_{cc})_i H^l_{jk}D^{ij} (\mathcal{B}_8)^{k}_l
   \nonumber\\
 & + b_4 (\mathcal{B}_{cc})_i H^l_{jk}D^{ik} (\mathcal{B}_8)^{j}_l + b_5 (\mathcal{B}_{cc})_i H^l_{jk}D^{jk} (\mathcal{B}_8)^{i}_l+ b_6 (\mathcal{B}_{cc})_i H^l_{jl}D^{ik} (\mathcal{B}_8)^{j}_k\nonumber\\
  &  + b_7 (\mathcal{B}_{cc})_i H^l_{jl}D^{jk} (\mathcal{B}_8)^{i}_k
   + b_{8} (\mathcal{B}_{cc})_i H^l_{lj}D^{ik} (\mathcal{B}_8)^{j}_k + b_{9} (\mathcal{B}_{cc})_i H^l_{lj}D^{jk} (\mathcal{B}_8)^{i}_k,
\end{align}
where $D^{ij} = \epsilon^{ijk}D_k$.
By inserting Eqs.~\eqref{sy1} and $D_k = \epsilon_{ijk}D^{ij}/2$ into each term of Eqs.~\eqref{a2} and \eqref{a3}, we derive the relations between decay amplitudes constructed by third-rank and $(1,1)$-rank octets,
\begin{align}
  b_1 & = \frac{1}{\sqrt{6}}(B^S_1+B^S_2-2B^S_3-B^S_5+B^S_6)
  +\frac{1}{\sqrt{2}}(-B^A_1-B^A_2+B^A_4),\nonumber\\
  b_2 & = \frac{1}{\sqrt{6}}(B^S_1-2B^S_2+B^S_3-B^S_4+B^S_6)
  +\frac{1}{\sqrt{2}}(B^A_1-B^A_3+B^A_5),\nonumber\\
 b_3 & = \frac{1}{\sqrt{6}}(-B^S_5+B^S_6)
  +\frac{1}{\sqrt{2}}B^A_4, \qquad b_4 = \frac{1}{\sqrt{6}}(-B^S_4+B^S_6)
  +\frac{1}{\sqrt{2}}B^A_5, \nonumber\\ b_5 & = \frac{1}{\sqrt{6}}(-B^S_4+B^S_5)
  +\frac{1}{\sqrt{2}}B^A_6,\nonumber\\
  b_6 & = \frac{1}{\sqrt{6}}(B^S_4-B^S_6+B^S_7+B^S_8-2B^S_9)
  +\frac{1}{\sqrt{2}}(-B^A_5-B^A_7-B^A_8),\nonumber\\
  b_7 & = \frac{1}{\sqrt{6}}(B^S_4-B^S_5+B^S_7-2B^S_8+B^S_9)
  +\frac{1}{\sqrt{2}}(-B^A_6+B^A_7-B^A_9),\nonumber\\
  b_8 & = \frac{1}{\sqrt{6}}(B^S_5-B^S_6+B^S_{10}+B^S_{11}-2B^S_{12})
  +\frac{1}{\sqrt{2}}(-B^A_4-B^A_{10}-B^A_{11}),\nonumber\\
  b_9 & = \frac{1}{\sqrt{6}}(-B^S_4+B^S_5+B^S_{10}-2B^S_{11}+B^S_{12})
  +\frac{1}{\sqrt{2}}(B^A_6+B^A_{10}-B^A_{12}).
\end{align}
The decay amplitudes constructed by (1,1)-rank octet tensors are listed in Table~\ref{ampb}.
Note that there are seven tree-induced amplitudes, $b_{1}\sim b_7$, and eight penguin-induced amplitudes, $b_1^P\sim b_3^P$ and $b_5^P\sim b_9^P$, contributing to the $\mathcal{B}_{cc}\to \mathcal{B}_{8}D$ decays.

The $SU(3)$ irreducible amplitude of $\mathcal{B}_{cc}\to \mathcal{B}_8 D$ can be expressed as
\begin{align}\label{b2}
  \mathcal{A}^{IR}(\mathcal{B}_{cc}\to \mathcal{B}_8 D) &=  b^\prime_1 (\mathcal{B}_{cc})_i H(15)^i_{jk}D^{jl} (\mathcal{B}_8)^{k}_l + b^\prime_2 (\mathcal{B}_{cc})_i H(15)^l_{jk}D^{ij} (\mathcal{B}_8)^{k}_l+b^\prime_3 (\mathcal{B}_{cc})_i H(\overline 6)^i_{jk}D^{jl} (\mathcal{B}_8)^{k}_l
   \nonumber\\
 & + b^\prime_4 (\mathcal{B}_{cc})_i H(\overline 6)^l_{jk}D^{ij} (\mathcal{B}_8)^{k}_l + b^\prime_5 (\mathcal{B}_{cc})_i H(\overline 6)^l_{jk}D^{jk} (\mathcal{B}_8)^{i}_l+ b^\prime_6 (\mathcal{B}_{cc})_i H(3)_{j}D^{ik} (\mathcal{B}_8)^{j}_k\nonumber\\
  &  + b^\prime_7 (\mathcal{B}_{cc})_i H(3)_{j}D^{jk} (\mathcal{B}_8)^{i}_k
   + b^\prime_{8} (\mathcal{B}_{cc})_i H(3^\prime)_{j}D^{ik} (\mathcal{B}_8)^{j}_k + b^\prime_{9} (\mathcal{B}_{cc})_i H(3^\prime)_{j}D^{jk} (\mathcal{B}_8)^{i}_k.
\end{align}
According to Eq.~\eqref{su}, the relations between the amplitudes of $\mathcal{B}_{cc}\to \mathcal{B}_8 D$ decays, as defined in Eq.~\eqref{b1} and Eq.~\eqref{b2}, are derived as
\begin{align}
   b_1^\prime & = b_1 + b_2,\qquad b_2^\prime = b_3 + b_4,\qquad b_3^\prime = b_1 - b_2,\qquad b_4^\prime = b_3 - b_4,\qquad b_5^\prime = b_5,\nonumber\\
b_6^\prime & = -\frac{1}{8}b_1 + \frac{3}{8}b_2 -\frac{1}{8}b_3 + \frac{3}{8}b_4 + b_6,\qquad b_7^\prime = \frac{3}{8}b_1 - \frac{1}{8}b_2 +\frac{1}{2}b_5 + b_7, \nonumber\\
b_8^\prime & = \frac{3}{8}b_1 - \frac{1}{8}b_2 +\frac{3}{8}b_3 - \frac{1}{8}b_4 + b_8,\qquad b_9^\prime = -\frac{1}{8}b_1 + \frac{3}{8}b_2 -\frac{1}{2}b_5 + b_9.
\end{align}
The three terms constructed by $H(\overline 6)_{ij}^k$ in Eq.~\eqref{b2} can be written as
\begin{align}\label{xp}
b^\prime_3 (\mathcal{B}_{cc})_i H(\overline 6)^i_{jk}D^{jl} (\mathcal{B}_8)^{k}_l
&= -b^\prime_3 (\mathcal{B}_{cc})_i H(\overline 6)^{il}D_{k} (\mathcal{B}_8)^{k}_l,\nonumber\\
b^\prime_4 (\mathcal{B}_{cc})_i H(\overline 6)^l_{jk}D^{ij} (\mathcal{B}_8)^{k}_l&=-b^\prime_4 (\mathcal{B}_{cc})_i H(\overline 6)^{kl}D_{k} (\mathcal{B}_8)^{i}_l+b^\prime_4 (\mathcal{B}_{cc})_i H(\overline 6)^{il}D_{k} (\mathcal{B}_8)^{k}_l,\nonumber\\ b^\prime_5 (\mathcal{B}_{cc})_i H(\overline 6)^l_{jk}D^{jk} (\mathcal{B}_8)^{i}_l& =2b^\prime_5(\mathcal{B}_{cc})_i H(\overline 6)^{kl}D_{k} (\mathcal{B}_8)^{i}_l.
\end{align}
Then we can define two new parameters to replace $b^\prime_3\sim b^\prime_5$,
\begin{align}\label{redef1}
b^{\prime\prime}_3 = b^{\prime}_3-b^{\prime}_4,\qquad b^{\prime\prime}_4 = b^{\prime}_4-2b^{\prime}_5.
\end{align}
Similarly to the $\mathcal{B}_{cc}\to \mathcal{B}_{c}M$ decays, the penguin-induced amplitudes appear together with the tree-induced amplitudes $b_6$ and $b_7$ in the following fixed combinations,
\begin{align}
  b^{ T+P}_6 = b_6 + b_1^P + b_3^P + b_6^P + 3b_8^P,\qquad  b^{ T+P}_7 = b_7 + b_2^P - b_5^P + b_7^P + 3b_9^P.
\end{align}
Ultimately, there are six independent amplitudes, $b_1^\prime, b_2^\prime, b_3^{\prime\prime}$, $b_4^{\prime\prime}$, $b_6^{T+P}$, and $b_7^{T+P}$, contributing to the $\mathcal{B}_{cc}\to \mathcal{B}_{8}D$ decays, and four of them dominate the branching fractions.

\begin{figure}
  \centering
  \includegraphics[width=7.5cm]{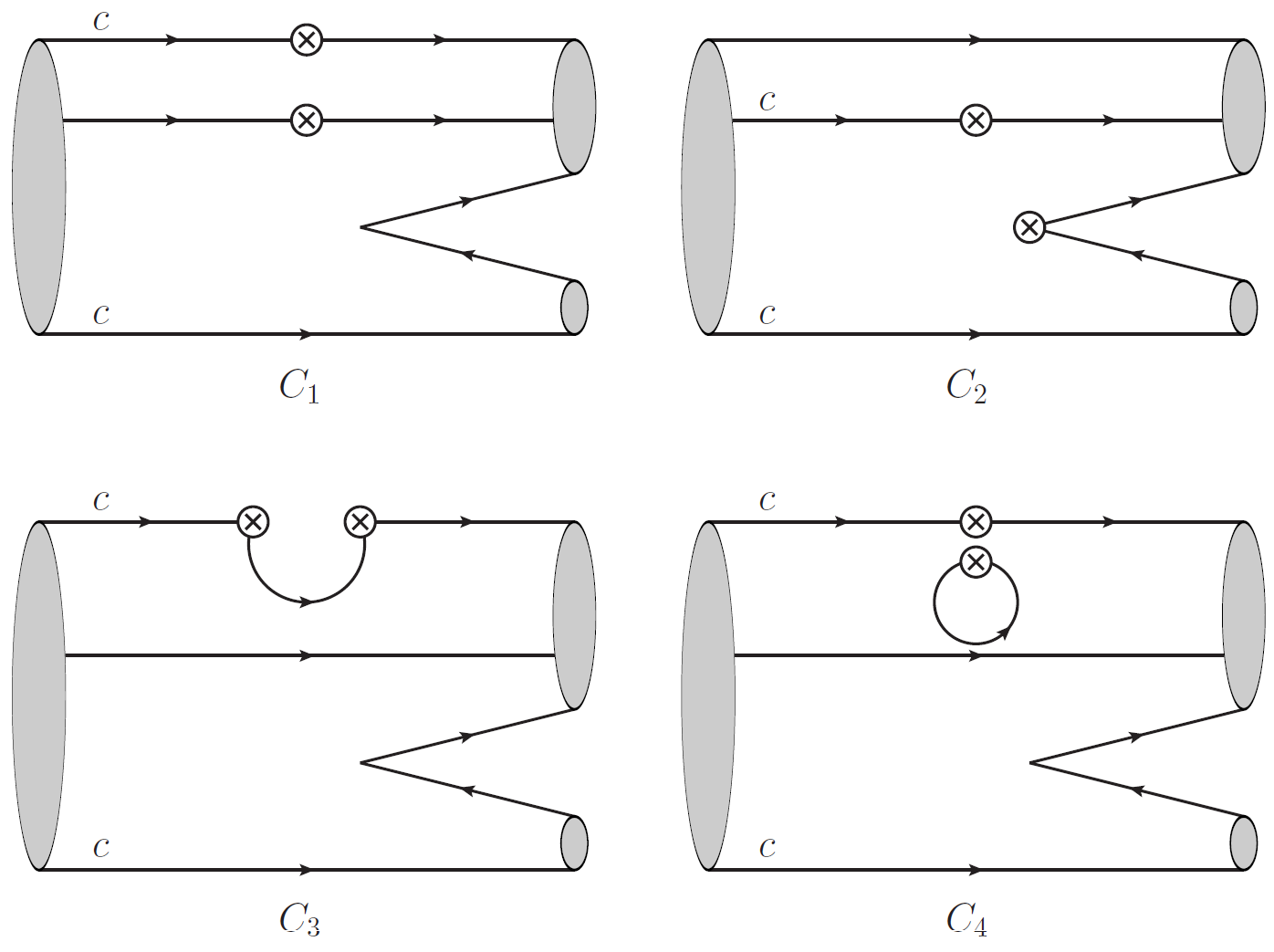}
  \caption{Topological diagrams in $\Xi_{cc}$ and $\Omega_{cc}$ decays into a charmed meson and a light decuplet baryon.}\label{top2}
\end{figure}

The light baryon decuplet is given as
\begin{align}
 &\Delta^{++}  = uuu,  \qquad \Delta^{-}=ddd,\qquad  \Omega^-=sss, \qquad\Sigma^{*0}=\frac{1}{\sqrt{6}}(uds +usd +dus+ dsu + sud +sdu),\nonumber\\
 & \Delta^{+} = \frac{1}{\sqrt{3}}(uud + udu +duu),\qquad  \Delta^{0} = \frac{1}{\sqrt{3}}(udd + dud +ddu),\qquad \Sigma^{*+}= \frac{1}{\sqrt{3}}(uus +usu +suu), \nonumber\\
&\Sigma^{*-}=\frac{1}{\sqrt{3}} (dds +dsd +sdd), \qquad
 \Xi^{*0}=\frac{1}{\sqrt{3}}(uss +sus+ssu),\qquad
 \Xi^{*-}=\frac{1}{\sqrt{3}}(dss + sds+ssd).
\end{align}
The amplitude of doubly charmed baryon decays into a charmed meson and a light decuplet baryon can be constructed as
\begin{align}
  \mathcal{A}(\mathcal{B}_{cc}\to \mathcal{B}_{10}D) = & C_1(\mathcal{B}_{cc})_iH^i_{jk}D_l\mathcal{B}_{10}^{jkl}+  C_2(\mathcal{B}_{cc})_iH^j_{kl}D_j\mathcal{B}_{10}^{ikl}+ C_3(\mathcal{B}_{cc})_iH^l_{jl}D_k\mathcal{B}_{10}^{ijk}
  +C_4(\mathcal{B}_{cc})_iH^l_{lj}D_k\mathcal{B}_{10}^{ijk}.
\end{align}
Since the three light quark indices in $\mathcal{B}_{10}^{ijk}$ are symmetric, the number of decay amplitudes contributing to the $\mathcal{B}_{cc}\to \mathcal{B}_{10}D$ mode is computed as $A^4_4/A^3_3 = 4$.
The topological diagrams in $\mathcal{B}_{cc}\to \mathcal{B}_{10}D$ decay are shown in Fig.~\ref{top2}, and the decay amplitudes of $\mathcal{B}_{cc}\to \mathcal{B}_{10}D$ are listed in Table~\ref{amp3}.
There are three tree-induced amplitudes, $C_{1}\sim C_3$, and four penguin-induced amplitudes, $C^P_{1}\sim C^P_{4}$, contributing to the $\mathcal{B}_{cc}\to \mathcal{B}_{10}D$ decays.

The $SU(3)$ irreducible amplitude of $\mathcal{B}_{cc}\to \mathcal{B}_{10}D$ decay is
\begin{align}
  \mathcal{A}^{IR}(\mathcal{B}_{cc}\to \mathcal{B}_{10}D) & =  c_1(\mathcal{B}_{cc})_iH(15)^i_{jk}D_l\mathcal{B}_{10}^{jkl}+  c_2(\mathcal{B}_{cc})_iH(15)^j_{kl}D_j\mathcal{B}_{10}^{ikl}+ c_3(\mathcal{B}_{cc})_iH(3)_{j}D_k\mathcal{B}_{10}^{ijk}\nonumber\\&
  +c_4(\mathcal{B}_{cc})_iH(3^\prime)_{j}D_k\mathcal{B}_{10}^{ijk}.
\end{align}
According to Eq.~\eqref{su}, the relations between the $SU(3)$ irreducible amplitudes and the topological amplitudes in $\mathcal{B}_{cc}\to \mathcal{B}_{10}D$ decay are derived to be
\begin{align}
   & c_1 = C_1,\qquad c_2 = C_2,  \qquad c_3 = \frac{1}{4}C_1 + \frac{1}{4}C_2 + C_3,  \qquad c_4 = \frac{1}{4}C_1 + \frac{1}{4}C_2 + C_4.
\end{align}
The penguin-induced diagrams appear together with tree-induced diagram $C_3$ in the fixed combination,
\begin{align}
  C_3^{T+P} &= C_3 + C^P_1 + C^P_2+ C^P_3 + 3C^P_4.
\end{align}
Thus, there are three independent amplitudes contributing to the $\mathcal{B}_{cc}\to \mathcal{B}_{10}D$ decays, and two of them dominate the branching fractions.

\section{Phenomenological analysis}\label{pa}

\subsection{Testing the flavor sum rules}

In Ref.~\cite{Luo:2023vbx}, we derive the isospin sum rules for singly and doubly charmed baryon decays by applying the operators $I_-^n$ on the initial and final states.
Since the topological amplitudes are expressed explicitly in this work, the isospin sum rules can be verified using the topological diagram approach.
For example, the isospin sum rule
\begin{align}
{ SumI_-}\,[\Xi^+_{cc}, \pi^+,\Xi^{+}_c]&= -\sqrt{2}\,\mathcal{A}(\Xi^+_{cc}\to \Xi^{+}_c\pi^0)- \mathcal{A}(\Xi^{++}_{cc}\to \Xi^{+}_c\pi^+)+\mathcal{A}(\Xi^+_{cc}\to \Xi^{0}_c\pi^+)=0
\end{align}
is verified by the topological amplitudes given in Table~\ref{amp1},
\begin{align}
\mathcal{A}(\Xi^{++}_{cc}\to \Xi^{+}_c\pi^+)& = \lambda_1(A_3+A_5),\qquad \mathcal{A}(\Xi^{+}_{cc}\to \Xi^{+}_c\pi^0) = -\frac{1}{\sqrt{2}}\lambda_1(A_1+A_3),\nonumber\\ \mathcal{A}(\Xi^+_{cc}\to \Xi^{0}_c\pi^+)&=\lambda_1(-A_1+A_5).
\end{align}
In Ref.~\cite{Wang:2022kwe}, we also derive the $U$-spin sum rules for the $\mathcal{B}_{cc} \to \mathcal{B}_{c\overline{3}} M$ and $\mathcal{B}_{cc} \to \mathcal{B}_{c6} M$ decays using the operator $S = -\lambda U_3 - \lambda^2 U_- + U_+$, which can also be verified using the topological amplitudes.
In this work, we supplemented the $U$-spin sum rules for the $\mathcal{B}_{cc}\to \mathcal{B}_{8}D$ decays in Appendix~\ref{U}.
For the $\mathcal{B}_{cc}\to \mathcal{B}_{10}D$ decays, only two topological amplitudes, $C_1$ and $C_2$, dominate.
There are rate sum rules that can be tested directly in experiments:
\begin{align}
  &|\mathcal{A}(\Xi_{cc}^{++}\to \Sigma^{*+}D^+)|^2 = |\mathcal{A}(\Omega_{cc}^{+}\to \Xi^{*0}D^+)|^2=\frac{|\mathcal{A}(\Xi_{cc}^{++}\to \Delta^{+}D^+)|^2}{\lambda^2} \nonumber\\&\qquad=\frac{|\mathcal{A}(\Xi_{cc}^{++}\to \Sigma^{*+}D^+_s)|^2}{\lambda^2}=\frac{|\mathcal{A}(\Xi_{cc}^{++}\to \Delta^{+} D^+_s)|^2}{\lambda^4}=\frac{|\mathcal{A}(\Xi_{cc}^{+}\to \Delta^{0} D^+_s)|^2}{\lambda^4},
\end{align}
\begin{align}
  &|\mathcal{A}(\Xi_{cc}^{+}\to \Sigma^{*+}D^0)|^2 =|\mathcal{A}(\Xi_{cc}^{+}\to \Xi^{*0}D^+_s)|^2=\frac{|\mathcal{A}(\Xi_{cc}^{+}\to \Delta^{+}D^0)|^2}{\lambda^2}\nonumber\\&\qquad=\frac{|\mathcal{A}(\Omega_{cc}^{+}\to \Sigma^{*+}D^0)|^2}{\lambda^2}=\frac{|\mathcal{A}(\Omega_{cc}^{+}\to \Delta^{+} D^0)|^2}{\lambda^4}=\frac{|\mathcal{A}(\Omega_{cc}^{+}\to \Delta^{0} D^+)|^2}{\lambda^4},
\end{align}
\begin{align}
  |\mathcal{A}(\Xi_{cc}^{+}\to \Sigma^{*0}D^+)|^2 =\frac{|\mathcal{A}(\Xi_{cc}^{+}\to \Delta^{0}D^+)|^2}{\lambda^2} = \frac{|\mathcal{A}(\Omega_{cc}^{+}\to \Xi^{*0}D^+_s)|^2}{\lambda^2} = \frac{|\mathcal{A}(\Omega_{cc}^{+}\to \Sigma^{*0} D^+_s)|^2}{\lambda^4},
\end{align}
\begin{align}
|\mathcal{A}(\Xi_{cc}^{+}\to \Sigma^{*0}D^+_s)|^2 = |\mathcal{A}(\Omega_{cc}^{+}\to \Sigma^{*0}D^+)|^2.
\end{align}

The decay amplitude for the $\mathcal{B}_{cc}\to\mathcal{B}M$ mode can be written as
\begin{eqnarray}\label{eq:A&B}
\mathcal{A}(\mathcal{B}_{cc}\to\mathcal{B}M)=i\overline u_\mathcal{B}(A+B\gamma_5)u_{\mathcal{B}_{cc}},
\end{eqnarray}
where $A$ and $B$ are the parity-violating $S$-wave and parity-conserving $P$-wave amplitudes with strong phases $\delta_S$ and $\delta_P$, respectively.
The flavor sum rules apply to all partial waves.
If two decay channels satisfy a flavor sum rule, their branching fractions are proportional, and their decay asymmetries $\alpha$, $\beta$, and $\gamma$ are identical.
Thus, we can use the flavor sum rules together with experimental data to predict the branching fractions and decay parameters of unobserved decay modes.
The decay width $\Gamma$ and decay parameters $\alpha$, $\beta$, and $\gamma$ can be calculated as
\begin{align}
&\Gamma = \frac{p_c}{8\pi}\frac{(m_{\mathcal{B}_{cc}}+m_\mathcal{B})^2-m_M^2}
{m_{\mathcal{B}_{cc}}^2}\left(|A|^2
+ \kappa^2|B|^2\right),
\nonumber\\
& \alpha=\frac{2\kappa |A^*B|\cos(\delta_P-\delta_S)}{|A|^2+\kappa^2 |B|^2},\quad
\beta=\frac{2\kappa |A^*B|\sin(\delta_P-\delta_S)}{|A|^2+\kappa^2 |B|^2},\quad
\gamma=\frac{|A|^2-\kappa^2 |B|^2}{|A|^2+\kappa^2 |B|^2},
\end{align}
where $p_c$ is the center-of-momentum (CM) in the rest frame of the initial baryon, and $\kappa$ is defined as $\kappa=p_c/(E_\mathcal{B}+m_\mathcal{B})=\sqrt{(E_\mathcal{B}-m_\mathcal{B})
/(E_\mathcal{B}+m_\mathcal{B})}$.
For doubly charmed baryon decays, only the branching fraction ratio $\mathcal{B}r(\Xi^{++}_{cc}\to \Xi^{* +}_c\pi^+)/\mathcal{B}r(\Xi^{++}_{cc}\to \Xi^{+}_c\pi^+)$ is measured as $1.41\pm 0.20$ \cite{LHCb:2022rpd}.
Using the $U$-spin relations
\begin{align}
 \frac{\mathcal{A}(\Xi^{++}_{cc}\to \Sigma^{ +}_cK^+)}{\mathcal{A}(\Xi^{++}_{cc}\to \Xi^{* +}_c\pi^+)}=\frac{V^*_{cd}V_{us}}{V^*_{cs}V_{ud}},\qquad \frac{\mathcal{A}(\Xi^{++}_{cc}\to \Lambda^{+}_cK^+)}{\mathcal{A}(\Xi^{++}_{cc}\to \Xi^{+}_c\pi^+)}=\frac{V^*_{cd}V_{us}}{V^*_{cs}V_{ud}},
\end{align}
we can estimate the branching fraction ratio $\mathcal{B}r(\Xi^{++}_{cc}\to \Sigma^{ +}_cK^+)/\mathcal{B}r(\Xi^{++}_{cc}\to \Lambda^{+}_cK^+)$ as
\begin{align}
\frac{\mathcal{B}r(\Xi^{++}_{cc}\to \Sigma^{ +}_cK^+)}{\mathcal{B}r(\Xi^{++}_{cc}\to \Lambda^{+}_cK^+)}=1.37\pm 0.19.
\end{align}

As analyzed in Refs.~\cite{Wang:2019dls,Zhang:2025jnw}, if two decay modes are $U$-spin conjugate, i.e., the interchanges of $s\leftrightarrow d$ and $\overline s\leftrightarrow -\overline d$ in all initial and final states, their $U$-spin amplitudes are related by the $d \leftrightarrow s$ interchange of the CKM matrix elements except for a possible minus sign.
This conclusion can be verified using the topological amplitudes listed in Tables~\ref{amp1}$\sim$\ref{ampb}.
For example, the $\Xi_{cc}^{++}\to \Lambda_c^+\pi^+$ and $\Xi_{cc}^{++}\to \Xi_c^+K^+$ modes are $U$-spin conjugate.
The decay amplitudes for the $\Xi_{cc}^{++}\to \Lambda_c^+\pi^+$ and $\Xi_{cc}^{++}\to \Xi_c^+K^+$ modes are given by
\begin{align}
  \mathcal{A}(\Xi_{cc}^{++}\to \Lambda_c^+\pi^+)&=\lambda_d(A_3+A_5+A_7+A_8)+\lambda_s(A_7+A_8)\nonumber\\
&\qquad-\lambda_b(A_1^P-A_3^P+A_4^P+A_6^P+A_7^P+A_8^P+3A_{10}^P+3A_{11}^P),\\
 \mathcal{A}(\Xi_{cc}^{++}\to \Xi_c^+K^+)&=\lambda_d(A_7+A_8)+\lambda_s(A_3+A_5+A_7+A_8)\nonumber\\
 &\qquad-\lambda_b(A_1^P-A_3^P+A_4^P+A_6^P+A_7^P+A_8^P+3A_{10}^P+3A_{11}^P).
\end{align}
Note that their decay amplitudes are related by the interchanges of the CKM matrix elements, $V^*_{cd}V_{ud}\leftrightarrow V^*_{cs}V_{us}$.
According to this relation, we obtain the $CP$ violation relation between two $U$-spin conjugate modes $\mathcal{B}_{cc}\to \mathcal{B} M$ and $\mathcal{B}^\prime_{cc}\to \mathcal{B}^\prime M^\prime$ as \cite{Wang:2019dls,Zhang:2025jnw}
\begin{align}
 \frac{A_{CP}^{\rm dir}(\mathcal{B}^\prime_{cc}\to \mathcal{B}^\prime M^\prime)}{A_{CP}^{\rm dir}(\mathcal{B}_{cc}\to \mathcal{B} M)}\simeq -\frac{\mathcal{B}r(\mathcal{B}_{cc}\to \mathcal{B} M)\cdot\tau(\mathcal{B}^\prime_{cc})}{\mathcal{B}r(\mathcal{B}^\prime_{cc}\to \mathcal{B}^\prime M^\prime)\cdot\tau(\mathcal{B}_{cc})}.
\end{align}
To facilitate subsequent research, the $U$-spin conjugate particles involved in the doubly charmed baryon decays are listed as follows:
\begin{align}
  &\Xi_{cc}^{++}\leftrightarrow  \Xi_{cc}^{++}, \qquad \Xi_{cc}^{+}\leftrightarrow  \Omega_{cc}^{+},\qquad \Xi_{c}^{+}\leftrightarrow  \Lambda_{c}^{+}, \qquad \Xi_{c}^{0}\leftrightarrow  \Xi_{c}^{0}, \qquad \Sigma_{c}^{++}\leftrightarrow  \Sigma_{c}^{++},\qquad\Sigma_{c}^{+}\leftrightarrow  \Xi_{c}^{*+},\nonumber\\
 &\Sigma_{c}^{0}\leftrightarrow  \Omega_{c}^{0}, \qquad \Xi_{c}^{*0}\leftrightarrow  \Xi_{c}^{*0}, \qquad \pi^+\leftrightarrow K^+, \qquad \pi^-\leftrightarrow K^-, \qquad K^0\leftrightarrow \overline K^0, \qquad D^+\leftrightarrow D^+_s, \nonumber\\
  & D^0\leftrightarrow D^0, \qquad p\leftrightarrow \Sigma^+, \qquad n\leftrightarrow \Xi^0, \qquad \Sigma^-\leftrightarrow \Xi^-, \qquad \Delta^{++}\leftrightarrow  \Delta^{++},\qquad \Delta^{+}\leftrightarrow  \Sigma^{*+}, \nonumber\\
&\Delta^{0}\leftrightarrow  \Xi^{*0}, \qquad \Delta^{-}\leftrightarrow  \Omega^{-},\qquad \Sigma^{*0}\leftrightarrow \Sigma^{*0},
\qquad \Sigma^{*-}\leftrightarrow  \Xi^{*-}.
\end{align}

\subsection{Topological diagram and rescattering dynamics }
\begin{figure}[t!]
  \centering
  \includegraphics[width=10cm]{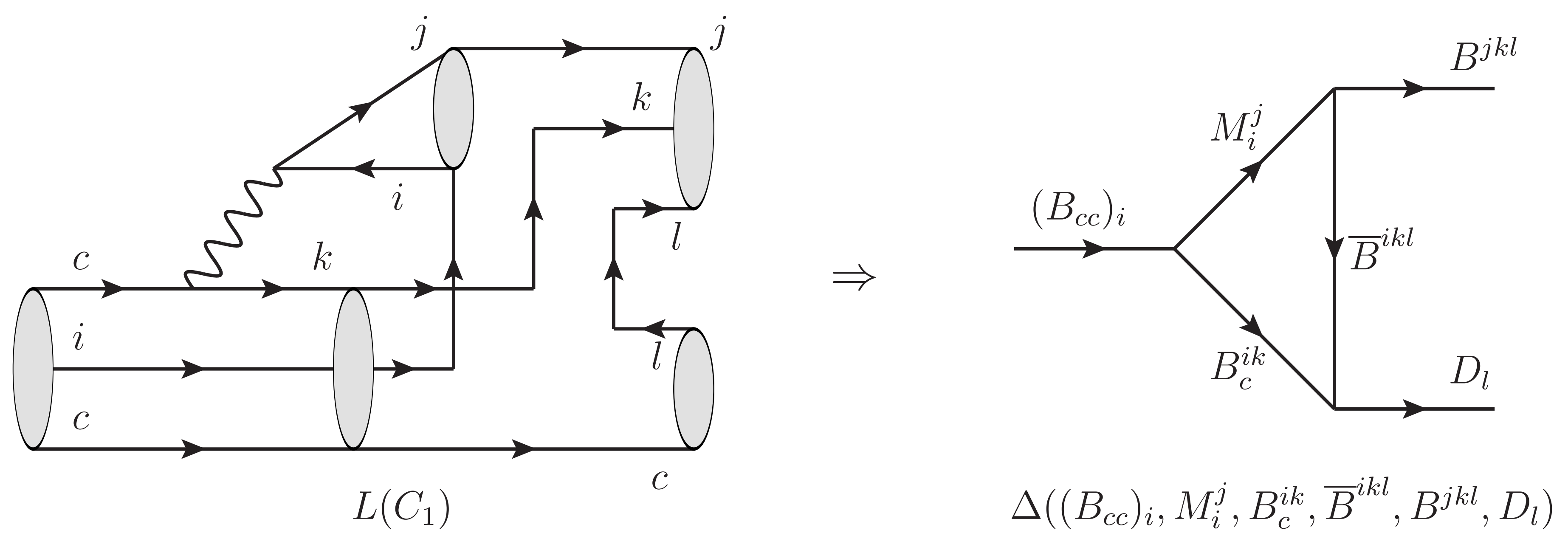}
  \caption{Long-distance rescattering contributions in diagram $C_1$.}\label{re3}
\end{figure}
\begin{figure}[t!]
  \centering
  \includegraphics[width=10cm]{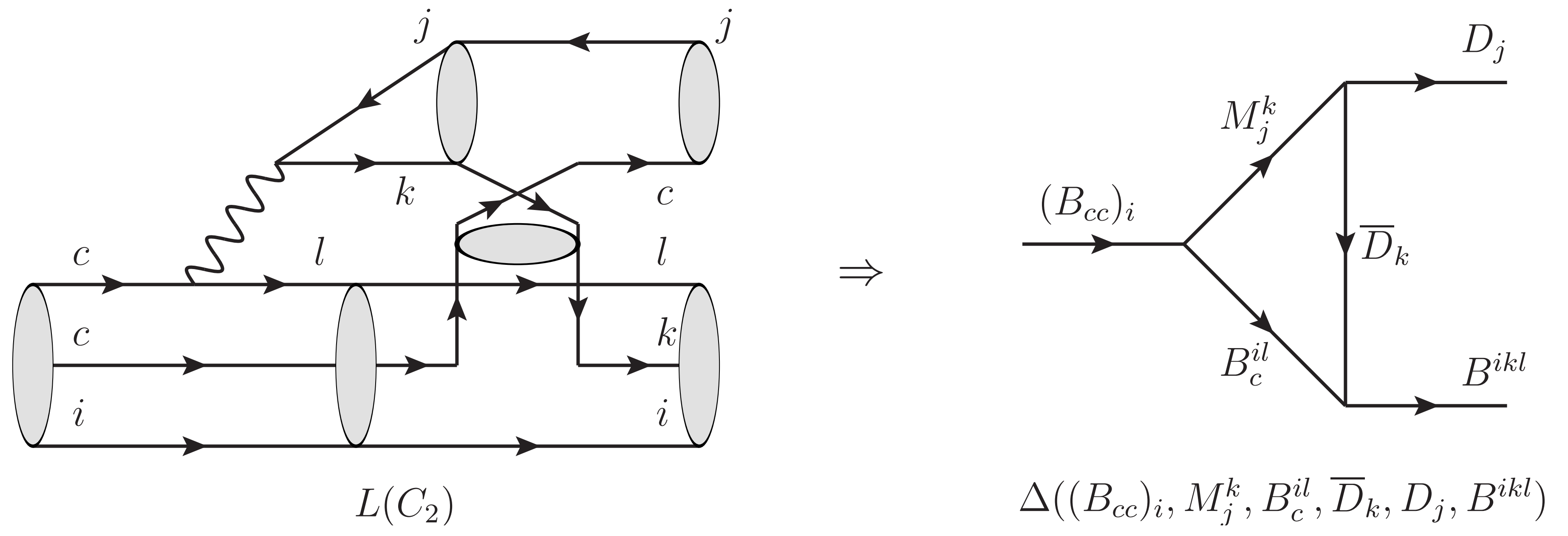}
  \caption{Long-distance rescattering contributions in diagram $C_2$.}\label{re4}
\end{figure}
\begin{figure}[t!]
  \centering
  \includegraphics[width=10cm]{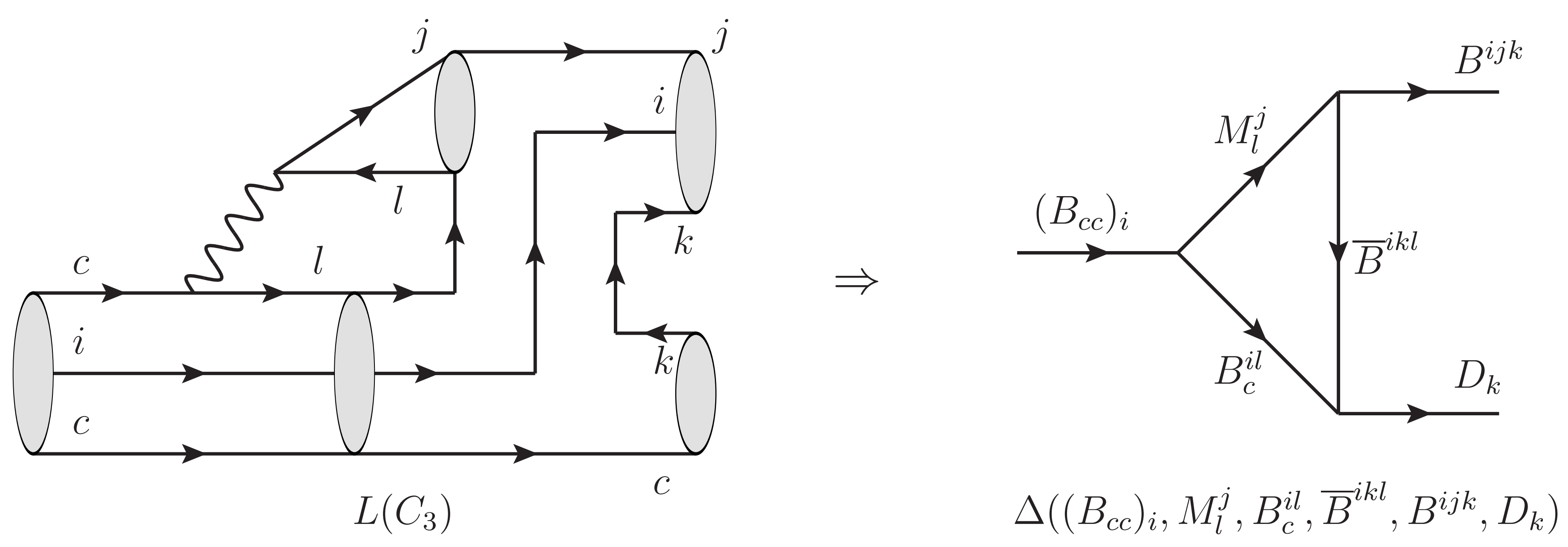}
  \caption{Long-distance rescattering contributions in diagram $C_3$.}\label{re5}
\end{figure}

For doubly charmed baryon decays into a charmed baryon and a light meson, the color-favored emitted diagrams $A_5$ and $A_5^{\prime}$ are the leading contributions in the $\mathcal{B}_{cc}\to \mathcal{B}_{c\overline 3}M$ and $\mathcal{B}_{cc}\to \mathcal{B}_{c6}M$ decays, respectively.
In the factorization approach, $A_5^{(\prime)}$ is dominated by the factorizable contribution $A_5^{(\prime)SD}$.
$A_5^{(\prime)SD}$ can be parameterized in terms of the decay constant of the emitted meson and the form factors of the baryon-to-baryon transition.
The factorizable part of the topological amplitude $A_4^{(\prime)}$ can be expressed as the Fierz transformation of $A_5^{(\prime)}$, which is suppressed by the small effective Wilson coefficient $a_2(m_c) = C_1(m_c) + C_2(m_c)/N_c \sim \mathcal{O}(0.1)$.
Thus, $A_4^{(\prime)SD}$ is typically predicted to be one order of magnitude smaller than $A_5^{(\prime)SD}$.

In addition to the factorizable contributions, there are significant non-factorizable contributions.
In the final-state interaction (FSI) framework, also referred to as rescattering dynamics, the non-factorizable quantum chromodynamics (QCD) effects can be modeled as the exchange of one particle between two particles generated from the tree-emitted amplitudes $A_5^{(\prime)SD}$.
There are $s$-channel and $t$-channel contributions in the rescattering dynamics, with the $t$-channel rescattering being dominant.
The $t$-channel rescattering forms a triangle diagram at the hadron level.
The calculations for charmed baryon decays in the framework of rescattering dynamics have been performed in Refs.~\cite{Yu:2017zst,Li:2020qrh,Hu:2024uia,Jiang:2018oak,Han:2021azw}.
The relation between topological diagrams and rescattering triangle diagrams was studied in Ref.~\cite{Wang:2021rhd} for charmed meson decays, and was extended to doubly charmed baryon decays into a charmed baryon and a light meson in Ref.~\cite{Wang:2022wrb}.
It was found that the triangle diagrams derived from the topological diagrams are consistent with those from the chiral Lagrangian.
By comparing Ref.~\cite{Wang:2022wrb} with Refs.~\cite{Hu:2024uia,Han:2021azw}, we can extract some useful knowledge about topological amplitudes as follows.
\begin{itemize}
  \item The rescattering contributions in the $\mathcal{B}_{cc}\to \mathcal{B}_{c\overline 3}M$ and $\mathcal{B}_{cc}\to \mathcal{B}_{c6}M$ decays are predicted to be smaller than the color-favored emitted diagrams, with $L(A_i^{(\prime)})/A_5^{(\prime)}\sim\mathcal{O}(0.1)$.
      The $\mathcal{B}_{cc}\to \mathcal{B}_{c\overline 3}M$ and $\mathcal{B}_{cc}\to \mathcal{B}_{c6}M$ decays are dominated by the emitted diagrams $A_5$ and $A_5^{\prime}$, respectively.
  \item There are two types of rescattering triangle diagrams contributing to doubly charmed baryon decays: two particles generated from a weak vertex scattering via the exchange of a meson or a baryon.
      They are predicted to be of the same order.
      Thus, the tree-induced diagrams in which the long-distance dynamics can be modeled as triangle diagrams, namely $A_1^{(\prime)}$, $A_3^{(\prime)}$, $A_4^{(\prime)}$, $A_6^{(\prime)}$, $A_7^{(\prime)}$, and $A_8^{(\prime)}$, are predicted to be of the same order.
      The diagrams in which the long-distance contributions cannot be modeled as triangle diagrams, $A_2^{(\prime)}$ and $A_9^{(\prime)}$, are considered smaller.
  \item Note that the diagrams $A_7^{(\prime)}$ and $A_8^{(\prime)}$ are quark-loop diagrams.
      The long-distance contributions in the quark-loop diagrams are of the same order as those in the tree diagrams.
      This indicates that the $CP$ violation in doubly charmed baryon decays could be of order $\mathcal{O}(10^{-4})$.
\end{itemize}

For doubly charmed baryon decays into a $D$ meson and a light baryon, the relations between the topological amplitudes and rescattering dynamics can be analyzed within the framework proposed in Ref.~\cite{Wang:2022wrb}.
The long-distance rescattering contributions in the $\mathcal{B}_{cc} \to \mathcal{B}_{10} D$ decays are shown in Figs.~\ref{re3}, \ref{re4}, and \ref{re5}.
For the $\mathcal{B}_{cc} \to \mathcal{B}_8 D$ decays, the long-distance rescattering contributions can be obtained by adding symmetric or antisymmetric marks in Figs.~\ref{re3}, \ref{re4}, and \ref{re5}.
For example, if the indices $j,k$ are marked as symmetric and antisymmetric in Fig.~\ref{re3}, they represent the long-distance rescattering contributions in the diagrams $B_1^S$ and $B_1^A$, respectively.
It was shown in Ref.~\cite{Li:2020qrh} that the branching fractions of the $\mathcal{B}_{cc}\to \mathcal{B}_8D$ decays are one order of magnitude smaller than those of the $\mathcal{B}_{cc}\to \mathcal{B}_cM$ modes dominated by $A_5$ or $A_5^{\prime}$.
This indicates that the tree-induced topological diagrams in the $\mathcal{B}_{cc}\to \mathcal{B}_8D$ decays are smaller than the color-favored emitted diagrams and of the same order as the non-factorizable diagrams in the $\mathcal{B}_{cc}\to \mathcal{B}_cM$ decays.

\subsection{Testing the K\"orner-Pati-Woo theorem}

In the literature, the K\"orner-Pati-Woo theorem \cite{Pati:1970fg,Korner:1970xq} is used to simplify the decay amplitudes of heavy baryon decays.
According to the K\"orner-Pati-Woo theorem, if the two quarks produced by weak operators enter one baryon, they must be antisymmetric in flavor.
For $\mathcal{B}_{cc}\to \mathcal{B}_c M$ decays, the two quark lines emitted from the weak vertex enter the final-state or resonance-state baryon in the tree-induced diagrams $A_1^{(\prime)}$, $A_2^{(\prime)}$, $A_3^{(\prime)}$, and  $A_6^{(\prime)}$.
Consequently, these quarks must be antisymmetric in flavor according to the K\"orner-Pati-Woo theorem.
The $\mathcal{O}(\overline{6})$ and $\mathcal{O}(15)$ operators are antisymmetric and symmetric, respectively, under the interchange of two covariant flavor indices.
Thus, only $\mathcal{O}(\overline 6)$ operators contribute to the diagrams $A_1^{(\prime)}$, $A_2^{(\prime)}$, $A_3^{(\prime)}$, and  $A_6^{(\prime)}$.
According to Eqs.~\eqref{sol} and \eqref{sol2}, we have
\begin{align}
 A_1 = -A_6,\qquad A^\prime_1 = -A^\prime_6, \qquad A^\prime_2=A^\prime_3=0.
\end{align}
Similarly, the K\"orner-Pati-Woo theorem results in the following relations for the diagrams contributing to the $\mathcal{B}_{cc}\to \mathcal{B}_8D$ and $\mathcal{B}_{cc}\to \mathcal{B}_{10}D$ decays,
\begin{align}
 B_2^{S,A} = B_3^{S,A},\qquad B_4^{S,A} = B_5^{S,A}, \qquad B^S_1=B^S_6=0
\end{align}
and
\begin{align}\label{kpw3}
 C_1 = C_2 = 0.
\end{align}
Neglecting the small $\lambda_b$ terms, all the $\mathcal{B}_{cc}\to \mathcal{B}_{10}D$ modes vanish under the $SU(3)_F$ limit.
Thus, the K\"orner-Pati-Woo theorem can be tested by measuring the branching fractions of any $\mathcal{B}_{cc}\to \mathcal{B}_{10}D$ modes.
In charmed hadron decays, the $SU(3)_F$ breaking effects are significant.
By combining isospin symmetry and the K\"orner-Pati-Woo theorem, we can derive several relations which go beyond the isospin sum rules given in Ref.~\cite{Luo:2023vbx}:
 \begin{align}\label{test1}
 \mathcal{A}(\Omega_{cc}^{+}\to \Sigma_c^{++}K^-) = \sqrt{2}\mathcal{A}(\Omega_{cc}^{+}\to \Sigma_c^{+} \overline K^0),
\end{align}
   \begin{align}
 \mathcal{A}(\Omega_{cc}^{+}\to \Sigma_c^{++}\pi^-) = -\mathcal{A}(\Omega_{cc}^{+}\to \Sigma_c^{+}\pi^0)=-\mathcal{A}(\Omega_{cc}^{+}\to \Sigma_c^{0}\pi^+),
\end{align}
 \begin{align}
 \mathcal{A}(\Xi_{cc}^{++}\to \Sigma_c^{++} K^0) = 2\mathcal{A}(\Xi_{cc}^{+}\to \Sigma_c^{+}K^0),
\end{align}
 \begin{align}
 2\mathcal{A}(\Xi_{cc}^{++}\to \Sigma_c^{+}K^+) = \mathcal{A}(\Xi_{cc}^{+}\to \Sigma_c^{0} K^+),
\end{align}
 \begin{align}
 \mathcal{A}(\Xi_{cc}^{++}\to \Sigma^{+}D^+_s) = -\sqrt{2}\mathcal{A}(\Xi_{cc}^{+}\to \Sigma^{0}D^+_s),
\end{align}
 \begin{align}
 \mathcal{A}(\Omega_{cc}^{+}\to \Sigma^{+}D^0) = -\sqrt{2}\mathcal{A}(\Omega_{cc}^{+}\to \Sigma^{0}D^+),
\end{align}
  \begin{align}
 \mathcal{A}(\Xi_{cc}^{++}\to p D^+_s) = \mathcal{A}(\Xi_{cc}^{+}\to n D^+_s),
\end{align}
 \begin{align}
 \mathcal{A}(\Omega_{cc}^{+}\to p D^0) = \mathcal{A}(\Omega_{cc}^{+}\to n D^+),
\end{align}
\begin{align}
 \mathcal{A}(\Xi^{++}_{cc}\to \Delta^+ D^+) = \mathcal{A}(\Xi^{+}_{cc}\to \Delta^+ D^0) = \mathcal{A}(\Xi^{+}_{cc}\to \Delta^0 D^+),
\end{align}
\begin{align}
 \mathcal{A}(\Xi^{++}_{cc}\to \Sigma^{*+} D^+_s) = \sqrt{2}\mathcal{A}(\Xi^{+}_{cc}\to \Sigma^{*0} D^+_s),
\end{align}
\begin{align}
 \mathcal{A}(\Omega^{+}_{cc}\to \Sigma^{*+} D^0) = \sqrt{2}\mathcal{A}(\Omega^{+}_{cc}\to \Sigma^{*0} D^+),
\end{align}
\begin{align}
 \mathcal{A}(\Xi^{++}_{cc}\to \Delta^{+} D^+_s) = \sqrt{2}\mathcal{A}(\Xi^{+}_{cc}\to \Delta^{0} D^+_s),
\end{align}
\begin{align}\label{test2}
 \mathcal{A}(\Omega^{+}_{cc}\to \Delta^{+} D^0) = \sqrt{2}\mathcal{A}(\Omega^{+}_{cc}\to \Delta^{0} D^+).
\end{align}
These relations provide an accurate test of the K\"orner-Pati-Woo theorem by measuring the branching fractions or decay parameters $\alpha$, $\beta$, and $\gamma$ in experiments.

\begin{figure}[t!]
  \centering
  \includegraphics[width=10cm]{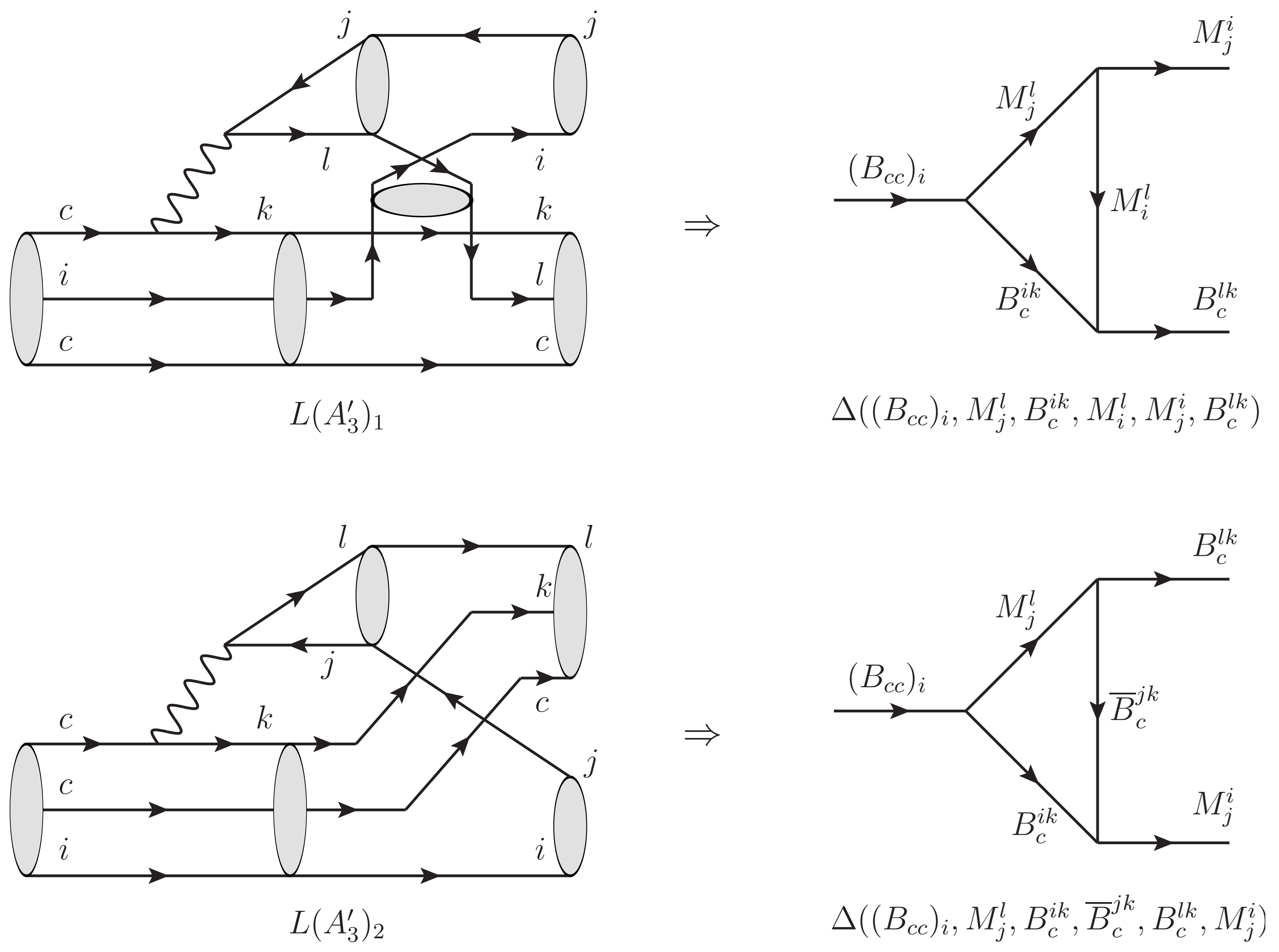}
  \caption{Long-distance rescattering contributions in diagram $A_3^\prime$, which is taken from Ref.~\cite{Wang:2022wrb}.}\label{re}
\end{figure}
\begin{figure}[t!]
  \centering
  \includegraphics[width=10cm]{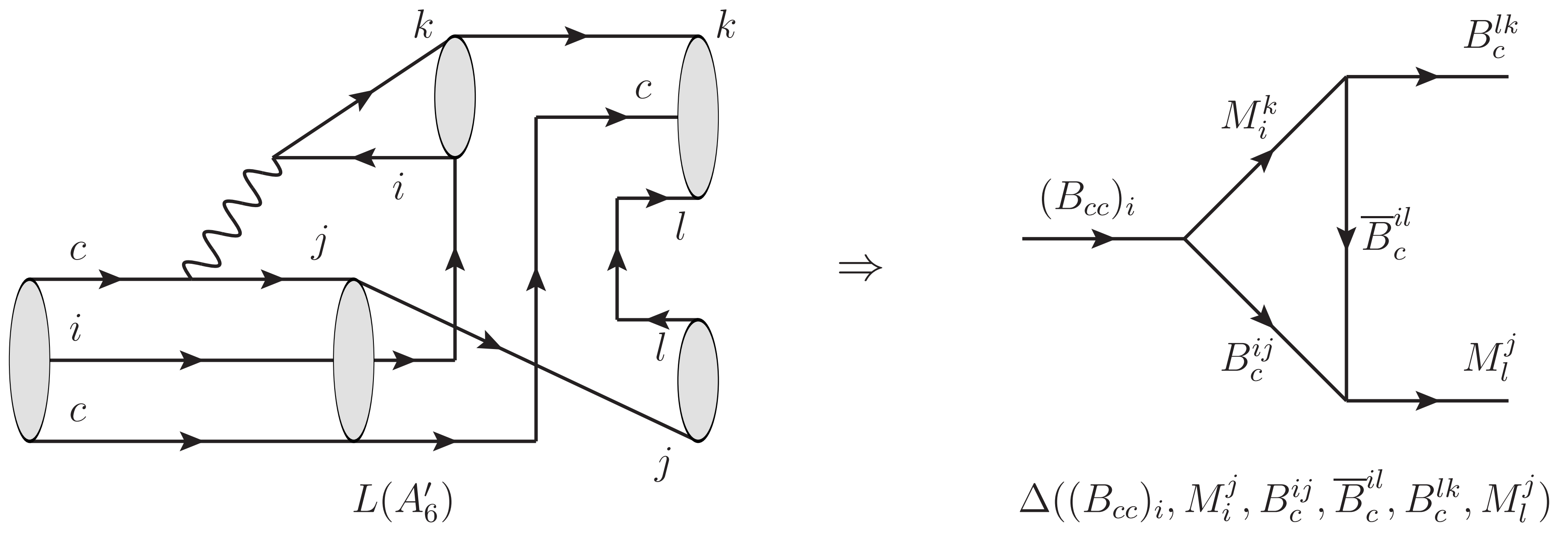}
  \caption{Long-distance rescattering contribution in diagram $A_6^\prime$, which is taken from Ref.~\cite{Wang:2022wrb}.}\label{re1}
\end{figure}
\begin{figure}[t!]
  \centering
  \includegraphics[width=10cm]{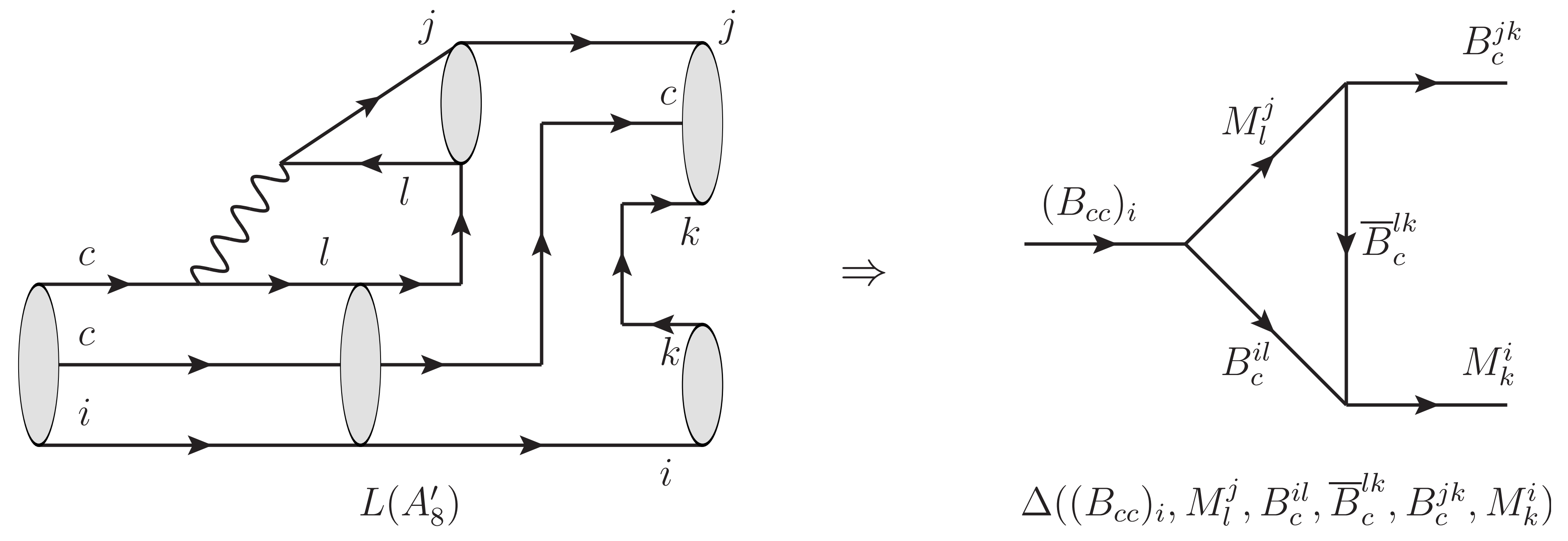}
  \caption{Long-distance rescattering contribution in diagram $A_8^\prime$, which is taken from Ref.~\cite{Wang:2022wrb}.}\label{re2}
\end{figure}

It was found in Ref.~\cite{Wang:2025khg} that the K\"orner-Pati-Woo theorem is inconsistent with the rescattering dynamics in singly charmed baryon decays.
This conflict also appears in doubly charmed baryon decays.
In doubly charmed baryon decays, Eqs.~\eqref{test1} to \eqref{test2} are violated in the rescattering triangle diagrams \cite{Li:2020qrh,Han:2021azw,Hu:2024uia}.
For example, the branching fraction of the $\Omega_{cc}^{+} \to \Sigma_c^{++} K^-$ mode should be twice that of the $\Omega_{cc}^{+} \to \Sigma_c^{+} \overline{K}^0$ mode if the K\"orner-Pati-Woo theorem holds.
Additionally, the decay parameters of the $\Omega_{cc}^{+} \to \Sigma_c^{++} K^-$ and $\Omega_{cc}^{+} \to \Sigma_c^{+} \overline{K}^0$ modes would be equivalent.
However, recent calculations based on rescattering dynamics show that \cite{Hu:2024uia}
\begin{align}
  \mathcal{B}r(\Omega_{cc}^{+}\to \Sigma_c^{++}K^-)&=(1.67^{+0.71+1.77}_{-0.35-1.01})\times 10^{-4},\qquad \mathcal{B}r(\Omega_{cc}^{+}\to \Sigma_c^{+}\overline K^0)=(2.09^{+1.15+2.14}_{-0.62-1.24})\times 10^{-4},\nonumber\\
  \alpha(\Omega_{cc}^{+}\to \Sigma_c^{++}K^-)&=0.33^{+0.18+0.01}_{-0.24-0.01},\qquad \alpha(\Omega_{cc}^{+}\to \Sigma_c^{+}\overline K^0)=-0.37^{+0.03+0.03}_{-0.07-0.03},\nonumber\\
  \beta(\Omega_{cc}^{+}\to \Sigma_c^{++}K^-)&=0.32^{+0.03+0.02}_{-0.07-0.02},\qquad \beta(\Omega_{cc}^{+}\to \Sigma_c^{+}\overline K^0)=-0.79^{+0.03+0.02}_{-0.02-0.02},\nonumber\\
  \gamma(\Omega_{cc}^{+}\to \Sigma_c^{++}K^-)&=0.89^{+0.04+0.00}_{-0.14-0.00},\qquad \gamma(\Omega_{cc}^{+}\to \Sigma_c^{+}\overline K^0)=-0.49^{+0.12+0.01}_{-0.06-0.01}.
\end{align}
These results are found to be inconsistent with Eq.~\eqref{test1}.
The reason is that the rescattering triangle diagrams for the $\Omega_{cc}^{+} \to \Sigma_c^{++} K^-$ and $\Omega_{cc}^{+} \to \Sigma_c^{+} \overline{K}^0$ modes are not proportional.
For the $\Omega_{cc}^{+} \to \Sigma_c^{++} K^-$ mode, only triangle diagrams with a baryon serving as the scattering propagator contribute.
For the $\Omega_{cc}^{+} \to \Sigma_c^{+} \overline{K}^0$ mode, both triangle diagrams with a baryon and those with a meson serving as the scattering propagator contribute.
See Refs.~\cite{Han:2021azw,Hu:2024uia} for details.
We can explain the difference between the $\Omega_{cc}^{+} \to \Sigma_c^{++} K^-$ and $\Omega_{cc}^{+} \to \Sigma_c^{+} \overline{K}^0$ modes from topological diagrams.
The tree-induced diagrams for the $\Omega_{cc}^{+} \to \Sigma_c^{++} K^-$ and $\Omega_{cc}^{+} \to \Sigma_c^{+} \overline{K}^0$ modes are
\begin{align}
  \mathcal{A}(\Omega_{cc}^{+}\to \Sigma_c^{++}K^-)&=\lambda_d A_8^{\prime}+\lambda_s(A_6^\prime+A_8^{\prime}),\qquad \mathcal{A}(\Omega_{cc}^{+}\to \Sigma_c^{+}\overline K^0)=\frac{1}{\sqrt{2}}\lambda_d(A_3^{\prime}+ A_8^{\prime})+\frac{1}{\sqrt{2}}\lambda_sA_8^{\prime}.
\end{align}
The long-distance rescattering contributions to the diagrams $A_3^{\prime}$, $A_6^{\prime}$, and $A_8^{\prime}$ are presented in Figs.~\ref{re}, \ref{re1}, and \ref{re2}, respectively.
It was found that the diagrams $A_6^{\prime}$ and $A_8^{\prime}$ receive contributions only from triangle diagrams with a baryon serving as the scattering propagator.
In contrast, the diagrams $A_3^{\prime}$ receive contributions from triangle diagrams with both a baryon and a meson serving as propagators.
This conflicts with the K\"orner-Pati-Woo theorem, as $A_3^\prime = 0$ if the K\"orner-Pati-Woo theorem holds.
To determine which theory is correct, definitive data from branching fractions or decay parameters are required.

\subsection{Topological diagram and large $N_c$ expansion}
\begin{figure}[t!]
  \centering
  \includegraphics[width=17cm]{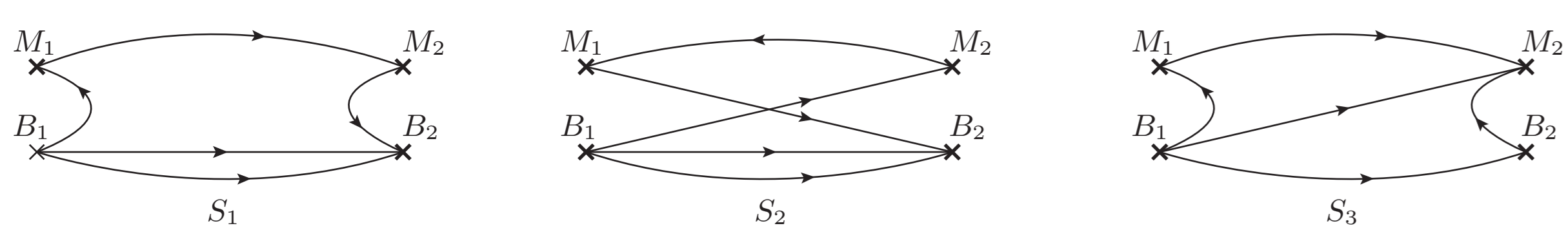}
  \caption{Sketches of meson-baryon scattering in the rescattering dynamics.}\label{re6}
\end{figure}

The large $N_c$ expansion is a useful theoretical framework for simplifying the study of strong interactions.
It treats the number of colors $N_c$ as a large parameter, expanding physical quantities in powers of $1/N_c$.
In the large $1/N_c$ limit, the 't Hooft coupling $\lambda = g^2 N_c$ is held fixed, leading to a dominance of planar diagrams while non-planar diagrams are suppressed \cite{tHooft:1973alw}.
Using the large $N_c$ expansion, we can roughly estimate the magnitude pattern of the topological diagrams.
The expansion orders for all tree-induced topological diagrams in doubly charmed baryon decays are listed as follows:
\begin{align}\label{s}
  \left\{
    \begin{array}{lll}
      A_5^{(\prime)} & \Rightarrow & (1/\sqrt{N_c})^3\times N_c^2 = \sqrt{N_c}; \\
      A_1^{(\prime)}, A_3^{(\prime)},A_4^{(\prime)},A_{6\sim8}^{(\prime)},B_{1\sim9}^{S,A}, C_{1\sim3} &\Rightarrow & (1/\sqrt{N_c})^3\times N_c = 1/\sqrt{N_c}; \\
      A_2^{(\prime)}, A_9^{(\prime)} & \Rightarrow& (1/\sqrt{N_c})^3\times N_c^2\times (1/N_c)^2 = (1/\sqrt{N_c})^{3}.
    \end{array}
  \right.
\end{align}

The magnitude pattern of topological diagrams in the large $N_c$ expansion is consistent with the predictions of rescattering dynamics given in Refs.~\cite{Yu:2017zst,Li:2020qrh,Hu:2024uia,Jiang:2018oak,Han:2021azw}.
This conformity can be explained by the following argument.
The sketches of meson-baryon scattering in the rescattering dynamics are shown in Fig.~\ref{re6}.
$S_1$ represents meson-baryon scattering induced by the insertion of two meson currents into one valence quark of the baryon.
It corresponds to the rescattering contributions in the diagrams $A_1^{(\prime)}$ and $A_7^{(\prime)}$.
$S_2$ represents meson-baryon scattering induced by a quark exchange between the meson and the baryon.
It corresponds to the rescattering contributions in the diagrams $A_3^{(\prime)}$, $A_4^{(\prime)}$, $B_{4 \sim 6}^{S,A}$, and $C_2$.
$S_3$ represents meson-baryon scattering induced by the insertion of two meson currents into two valence quarks of the baryon.
It corresponds to the rescattering contributions in the diagrams $A_6^{(\prime)}$, $A_8^{(\prime)}$, $B_{1 \sim 3}^{S,A}$, $B_{7 \sim 9}^{S,A}$, $C_1$, and $C_3$.
All $S_1 \sim S_3$ are of order $(1/\sqrt{N_c})^4 \times N_c = 1/N_c$ in the large $N_c$ expansion.
Then the order of the diagrams in which the long-distance contributions can be modeled as triangle diagrams is $\sqrt{N_c} \times 1/N_c = 1/\sqrt{N_c}$, which is consistent with Eq.~\eqref{s}.

\section{Summary}\label{summary}

In this work, we study the topological amplitudes of doubly charmed baryon decays in the $SU(3)_F$ limit.
The tree- and penguin-induced topological diagrams for the $\mathcal{B}_{cc}\to \mathcal{B}_{c\overline 3}M$, $\mathcal{B}_{cc}\to \mathcal{B}_{c 6}M$, $\mathcal{B}_{cc}\to \mathcal{B}_{8}D$, and $\mathcal{B}_{cc}\to \mathcal{B}_{10}D$ decays are presented.
The linear relations between the topological amplitudes and the $SU(3)$ irreducible amplitudes are derived through tensor contraction and $SU(3)$ decomposition.
The magnitude pattern of topological diagrams is analyzed within the frameworks of rescattering dynamics and large $N_c$ expansion.
The K\"orner-Pati-Woo theorem leads to several equations under isospin limits, which can be used to test its implication.

\begin{acknowledgements}

This work was supported in part by the National Natural Science Foundation of China under Grants No. 12105099.

\begin{appendix}
\section{$U$-spin sum rules for $\mathcal{B}_{cc}\to \mathcal{B}_{8}D$ decays}\label{U}

The $U$-spin sum rules for the $\mathcal{B}_{cc}\to \mathcal{B}_{8}D$ decays are listed as follows.
\begin{align}
SumS[\Xi_{cc}^{++},\Sigma^+,D^+]=-\frac{\mathcal{A}(\Xi^{++}_{cc}\to pD^+)}{\lambda} -\frac{\mathcal{A}(\Xi^{++}_{cc}\to \Sigma^+D^+_s)}{\lambda}+2\mathcal{A}(\Xi^{++}_{cc}\to \Sigma^+D^+)= 0,
\end{align}
\begin{align}
SumS[\Xi_{cc}^{++},p,D^+]=\mathcal{A}(\Xi^{++}_{cc}\to \Sigma^+D^+) -\frac{\mathcal{A}(\Xi^{++}_{cc}\to pD^+_s)}{\lambda^2}= 0,
\end{align}
\begin{align}
SumS[\Xi_{cc}^{++},p,D^+_s]=\mathcal{A}(\Xi^{++}_{cc}\to pD^+) +\mathcal{A}(\Xi^{++}_{cc}\to \Sigma^+D^+_s)-\frac{2\mathcal{A}(\Xi^{++}_{cc}\to pD^+_s)}{\lambda}= 0,
\end{align}
\begin{align}
SumS[\Xi_{cc}^{+},\Sigma^+,D^0]=-\frac{\mathcal{A}(\Xi^{+}_{cc}\to pD^0)}{\lambda} -\frac{\mathcal{A}(\Omega^{+}_{cc}\to \Sigma^+D^0)}{\lambda}+2\mathcal{A}(\Xi^{+}_{cc}\to \Sigma^+D^0)= 0,
\end{align}
\begin{align}
SumS[\Xi_{cc}^{+},p,D^0]=\mathcal{A}(\Xi^{+}_{cc}\to \Sigma^+D^0) -\frac{\mathcal{A}(\Omega^{+}_{cc}\to pD^0)}{\lambda^2}= 0,
\end{align}
\begin{align}
SumS[\Omega_{cc}^{+},p,D^0]=\mathcal{A}(\Xi^{+}_{cc}\to pD^0) +\mathcal{A}(\Omega^{+}_{cc}\to \Sigma^+D^0)-\frac{2\mathcal{A}(\Omega^{+}_{cc}\to pD^0)}{\lambda}= 0,
\end{align}
\begin{align}
SumS[\Xi_{cc}^{+},\Sigma^0,D^+]=\frac{\mathcal{A}(\Xi^{+}_{cc}\to nD^+)}{\sqrt{2}\lambda} -\frac{\mathcal{A}(\Xi^{+}_{cc}\to \Sigma^0D^+_s)}{\lambda}-\frac{\mathcal{A}(\Omega^{+}_{cc}\to \Sigma^0D^+)}{\lambda}+2\mathcal{A}(\Xi^{+}_{cc}\to \Sigma^0D^+)= 0,
\end{align}
\begin{align}
SumS[\Xi_{cc}^{+},n,D^+]=\frac{\sqrt{6}}{2}\mathcal{A}(\Xi^{+}_{cc}\to \Lambda^0D^+) -\frac{\mathcal{A}(\Xi^{+}_{cc}\to \Sigma^0D^+)}{\sqrt{2}}-\frac{\mathcal{A}(\Xi^{+}_{cc}\to nD^+_s)}{\lambda^2}-\frac{\mathcal{A}(\Omega^{+}_{cc}\to nD^+)}{\lambda^2}= 0,
\end{align}
\begin{align}
SumS[\Xi_{cc}^{+},\Xi^0,D^+]=\frac{\sqrt{6}}{2}\mathcal{A}(\Xi^{+}_{cc}\to \Lambda^0D^+) -\mathcal{A}(\Xi^{+}_{cc}\to \Xi^0D^+_s)-\frac{\mathcal{A}(\Xi^{+}_{cc}\to \Sigma^0D^+)}{\sqrt{2}}-\mathcal{A}(\Omega^{+}_{cc}\to \Xi^0D^+)= 0,
\end{align}
\begin{align}
SumS[\Xi_{cc}^{+},\Lambda^0,D^+]=-\frac{\sqrt{6}\mathcal{A}(\Xi^{+}_{cc}\to nD^+)}{2\lambda} -\frac{\mathcal{A}(\Xi^{+}_{cc}\to \Lambda^0D^+_s)}{\lambda}-\frac{\mathcal{A}(\Omega^{+}_{cc}\to \Lambda^0D^+)}{\lambda}+2\mathcal{A}(\Xi^{+}_{cc}\to \Lambda^0D^+)= 0,
\end{align}
\begin{align}
SumS[\Xi_{cc}^{+},\Sigma^0,D^+_s]=\frac{\mathcal{A}(\Xi^{+}_{cc}\to \Xi^0D^+_s)}{\sqrt{2}}+\mathcal{A}(\Xi^{+}_{cc}\to \Sigma^0D^+)+\frac{\mathcal{A}(\Xi^{+}_{cc}\to nD^+_s)}{\sqrt{2}\lambda^2}-\frac{\mathcal{A}(\Omega^{+}_{cc}\to \Sigma^0D^+_s)}{\lambda^2}= 0,
\end{align}
\begin{align}
SumS[\Xi_{cc}^{+},n,D^+_s]=\mathcal{A}(\Xi^{+}_{cc}\to nD^+)+\frac{\sqrt{6}}{2}\mathcal{A}(\Xi^{+}_{cc}\to \Lambda^0D^+_s)-\frac{\mathcal{A}(\Xi^{+}_{cc}\to \Sigma^0D^+_s)}{\sqrt{2}}-\frac{2\mathcal{A}(\Xi^{+}_{cc}\to nD^+_s)}{\lambda}= 0,
\end{align}
\begin{align}
SumS[\Xi_{cc}^{+},\Xi^0,D^+_s]=\frac{\sqrt{6}\mathcal{A}(\Xi^{+}_{cc}\to \Lambda^0D^+_s)}{2\lambda}-\frac{\mathcal{A}(\Xi^{+}_{cc}\to \Sigma^0D^+_s)}{\sqrt{2}\lambda}-\frac{\mathcal{A}(\Omega^{+}_{cc}\to \Xi^0D^+_s)}{\lambda}+2\mathcal{A}(\Xi^{+}_{cc}\to \Xi^0D^+_s)= 0,
\end{align}
\begin{align}
SumS[\Xi_{cc}^{+},\Lambda^0,D^+_s]=\mathcal{A}(\Xi^{+}_{cc}\to \Lambda^0D^+)-\frac{\sqrt{6}}{2}\mathcal{A}(\Xi^{+}_{cc}\to \Xi^0D^+_s)-\frac{\sqrt{6}\mathcal{A}(\Xi^{+}_{cc}\to nD^+_s)}{2\lambda^2}-\frac{\mathcal{A}(\Omega^{+}_{cc}\to \Lambda^0D^+_s)}{\lambda^2}= 0,
\end{align}
\begin{align}
SumS[\Omega_{cc}^{+},\Sigma^0,D^+]=\mathcal{A}(\Xi^{+}_{cc}\to \Sigma^0D^+)+\frac{\mathcal{A}(\Omega^{+}_{cc}\to \Xi^0D^+)}{\sqrt{2}}+\frac{\mathcal{A}(\Omega^{+}_{cc}\to nD^+)}{\sqrt{2}\lambda^2}-\frac{\mathcal{A}(\Omega^{+}_{cc}\to \Sigma^0D^+_s)}{\lambda^2}= 0,
\end{align}
\begin{align}
SumS[\Omega_{cc}^{+},n,D^+]=\mathcal{A}(\Xi^{+}_{cc}\to nD^+)+\frac{\sqrt{6}}{2}\mathcal{A}(\Omega^{+}_{cc}\to \Lambda^0D^+)-\frac{\mathcal{A}(\Omega^{+}_{cc}\to \Sigma^0D^+)}{\sqrt{2}}-\frac{2\mathcal{A}(\Omega^{+}_{cc}\to nD^+)}{\lambda}= 0,
\end{align}
\begin{align}
SumS[\Omega_{cc}^{+},\Xi^0,D^+]=\frac{\sqrt{6}\mathcal{A}(\Omega^{+}_{cc}\to \Lambda^0D^+)}{2\lambda}-\frac{\mathcal{A}(\Omega^{+}_{cc}\to \Xi^0D^+_s)}{\lambda}-\frac{\mathcal{A}(\Omega^{+}_{cc}\to \Sigma^0D^+)}{\sqrt{2}\lambda}+2\mathcal{A}(\Omega^{+}_{cc}\to \Xi^0D^+)= 0,
\end{align}
\begin{align}
SumS[\Omega_{cc}^{+},\Lambda^0,D^+]=\mathcal{A}(\Xi^{+}_{cc}\to \Lambda^0D^+)-\frac{\sqrt{6}}{2}\mathcal{A}(\Omega^{+}_{cc}\to \Xi^0D^+)-\frac{\sqrt{6}\mathcal{A}(\Omega^{+}_{cc}\to nD^+)}{2\lambda}+\frac{\mathcal{A}(\Omega^{+}_{cc}\to \Lambda^0D^+_s)}{\lambda^2}= 0,
\end{align}
\begin{align}
SumS[\Omega_{cc}^{+},\Sigma^0,D^+_s]=\mathcal{A}(\Xi^{+}_{cc}\to \Sigma^0D^+_s)+\frac{\mathcal{A}(\Omega^{+}_{cc}\to \Xi^0D^+_s)}{\sqrt{2}}+\mathcal{A}(\Omega^{+}_{cc}\to \Sigma^0D^+)-\frac{2\mathcal{A}(\Omega^{+}_{cc}\to \Sigma^0D^+_s)}{\lambda}= 0,
\end{align}
\begin{align}
SumS[\Omega_{cc}^{+},n,D^+_s]=\mathcal{A}(\Xi^{+}_{cc}\to nD^+_s)+\mathcal{A}(\Omega^{+}_{cc}\to nD^+)+\frac{\sqrt{6}}{2}\mathcal{A}(\Omega^{+}_{cc}\to \Lambda^0D^+_s)-\frac{\mathcal{A}(\Omega^{+}_{cc}\to \Sigma^0D^+_s)}{\sqrt{2}}= 0,
\end{align}
\begin{align}
SumS[\Omega_{cc}^{+},\Xi^0,D^+_s]=\mathcal{A}(\Xi^{+}_{cc}\to \Xi^0D^+_s)+\mathcal{A}(\Omega^{+}_{cc}\to \Xi^0D^+)+\frac{\sqrt{6}}{2}\mathcal{A}(\Omega^{+}_{cc}\to \Lambda^0D^+_s)-\frac{\mathcal{A}(\Omega^{+}_{cc}\to \Sigma^0D^+_s)}{\sqrt{2}}= 0,
\end{align}
\begin{align}
SumS[\Omega_{cc}^{+},\Lambda^0,D^+_s]=\mathcal{A}(\Xi^{+}_{cc}\to \Lambda^0D^+_s)+\mathcal{A}(\Omega^{+}_{cc}\to \Lambda^0D^+)-\frac{\sqrt{6}}{2}\mathcal{A}(\Omega^{+}_{cc}\to \Xi^0D^+_s)-\frac{2\mathcal{A}(\Omega^{+}_{cc}\to \Lambda^0D^+_s)}{\lambda}= 0.
\end{align}

\end{appendix}

\end{acknowledgements}


\begin{thebibliography}{99}

\bibitem{Aaij:2017ueg}
  R.~Aaij {\it et al.} [LHCb],
  Phys.\ Rev.\ Lett.\  {\bf 119}, no. 11, 112001 (2017)
  [arXiv:1707.01621 [hep-ex]].

\bibitem{Aaij:2018wzf}
  R.~Aaij {\it et al.} [LHCb],
  Phys.\ Rev.\ Lett.\  {\bf 121}, no. 5, 052002 (2018)
  [arXiv:1806.02744 [hep-ex]].


\bibitem{Aaij:2018gfl}
R.~Aaij \textit{et al.} [LHCb],
Phys. Rev. Lett. \textbf{121}, no.16, 162002 (2018)
[arXiv:1807.01919 [hep-ex]].

\bibitem{LHCb:2019qed}
R.~Aaij \textit{et al.} [LHCb],
Chin. Phys. C \textbf{44}, no.2, 022001 (2020)
[arXiv:1910.11316 [hep-ex]].

\bibitem{LHCb:2019ybf}
R.~Aaij \textit{et al.} [LHCb],
JHEP \textbf{10}, 124 (2019)
[arXiv:1905.02421 [hep-ex]].

\bibitem{LHCb:2022rpd}
R.~Aaij \textit{et al.} [LHCb],
JHEP \textbf{05}, 038 (2022)
[arXiv:2202.05648 [hep-ex]].

\bibitem{LHCb:2025shu}
R.~Aaij \textit{et al.} [LHCb],
JHEP \textbf{10}, 136 (2025)
[arXiv:2504.05063 [hep-ex]].

\bibitem{LHCb:2019gqy}
R.~Aaij \textit{et al.} [LHCb],
Sci. China Phys. Mech. Astron. \textbf{63}, no.2, 221062 (2020)
[arXiv:1909.12273 [hep-ex]].

\bibitem{LHCb:2020iko}
R.~Aaij \textit{et al.} [LHCb],
JHEP \textbf{11}, 095 (2020)
[arXiv:2009.02481 [hep-ex]].

\bibitem{LHCb:2021xba}
R.~Aaij \textit{et al.} [LHCb],
Chin. Phys. C \textbf{45}, no.9, 093002 (2021)
[arXiv:2104.04759 [hep-ex]].

\bibitem{LHCb:2021rkb}
R.~Aaij \textit{et al.} [LHCb],
Sci. China Phys. Mech. Astron. \textbf{64}, no.10, 101062 (2021)
[arXiv:2105.06841 [hep-ex]].

\bibitem{LHCb:2021eaf}
R.~Aaij \textit{et al.} [LHCb],
JHEP \textbf{12}, 107 (2021)
[arXiv:2109.07292 [hep-ex]].

\bibitem{LHCb:2022fbu}
R.~Aaij \textit{et al.} [LHCb],
Chin. Phys. C \textbf{47}, no.9, 093001 (2023)
[arXiv:2204.09541 [hep-ex]].

\bibitem{Yu:2017zst}
F.~S.~Yu, H.~Y.~Jiang, R.~H.~Li, C.~D.~L\"u, W.~Wang and Z.~X.~Zhao,
Chin. Phys. C \textbf{42}, no.5, 051001 (2018)
[arXiv:1703.09086 [hep-ph]].

\bibitem{Jiang:2018oak}
L.~J.~Jiang, B.~He and R.~H.~Li,
Eur. Phys. J. C \textbf{78}, no.11, 961 (2018)
[arXiv:1810.00541 [hep-ph]].

\bibitem{Li:2020qrh}
R.~H.~Li, J.~J.~Hou, B.~He and Y.~R.~Wang,
Chin. Phys. C \textbf{45}, no.4, 043108 (2021)
[arXiv:2010.09362 [hep-ph]].

\bibitem{Han:2021azw}
J.~J.~Han, H.~Y.~Jiang, W.~Liu, Z.~J.~Xiao and F.~S.~Yu,
Chin. Phys. C \textbf{45}, no.5, 053105 (2021)
[arXiv:2101.12019 [hep-ph]].

\bibitem{Hu:2024uia}
X.~H.~Hu, C.~P.~Jia, Y.~Xing and F.~S.~Yu,
Phys. Rev. D \textbf{111}, no.7, 076002 (2025)
[arXiv:2403.09511 [hep-ph]].

\bibitem{Wang:2017azm}
  W.~Wang, Z.~P.~Xing and J.~Xu,
  Eur.\ Phys.\ J.\ C {\bf 77}, no. 11, 800 (2017)
  [arXiv:1707.06570 [hep-ph]].

\bibitem{Shi:2017dto}
  Y.~J.~Shi, W.~Wang, Y.~Xing and J.~Xu,
  Eur.\ Phys.\ J.\ C {\bf 78}, no. 1, 56 (2018)
  [arXiv:1712.03830 [hep-ph]].

\bibitem{Li:2021rfj}
D.~M.~Li, X.~R.~Zhang, Y.~Xing and J.~Xu,
Eur. Phys. J. Plus \textbf{136}, no.7, 772 (2021)
[arXiv:2101.12574 [hep-ph]].

\bibitem{Wang:2024ztg}
D.~Wang and J.~F.~Luo,
Phys. Rev. D \textbf{110}, no.9, 093001 (2024)
[arXiv:2406.14061 [hep-ph]].

\bibitem{Wang:2024nxb}
D.~Wang,
Phys. Lett. B \textbf{858}, 139039 (2024)
[arXiv:2408.02015 [hep-ph]].

\bibitem{Wang:2025bdl}
D.~Wang and W.~C.~Fu,
Chin. Phys. \textbf{49}, no.11, 113104 (2025)
[arXiv:2503.22170 [hep-ph]].

\bibitem{Korner:1970xq}
J.~G.~K\"orner,
Nucl. Phys. B \textbf{25}, 282-290 (1971).

\bibitem{Pati:1970fg}
J.~C.~Pati and C.~H.~Woo,
Phys. Rev. D \textbf{3}, 2920-2922 (1971).

\bibitem{Jia:2024pyb}
C.~P.~Jia, H.~Y.~Jiang, J.~P.~Wang and F.~S.~Yu,
JHEP \textbf{11}, 072 (2024)
[arXiv:2408.14959 [hep-ph]].

\bibitem{Wang:2025khg}
D.~Wang,
[arXiv:2507.06914 [hep-ph]].

\bibitem{Buchalla:1995vs}
  G.~Buchalla, A.~J.~Buras and M.~E.~Lautenbacher,
  Rev.\ Mod.\ Phys.\  {\bf 68}, 1125 (1996)
  [hep-ph/9512380].

\bibitem{Beneke:1999br}
  M.~Beneke, G.~Buchalla, M.~Neubert and C.~T.~Sachrajda,
  Phys.\ Rev.\ Lett.\  {\bf 83}, 1914 (1999)
  [hep-ph/9905312].

\bibitem{Wang:2020gmn}
D.~Wang, C.~P.~Jia and F.~S.~Yu,
JHEP \textbf{09}, 126 (2021)
[arXiv:2001.09460 [hep-ph]].

\bibitem{Wang:2022kwe}
D.~Wang,
JHEP \textbf{12}, 003 (2022)
[arXiv:2204.05915 [hep-ph]].

\bibitem{Luo:2023vbx}
J.~F.~Luo and D.~Wang,
Eur. Phys. J. C \textbf{84}, no.10, 1010 (2024)
[arXiv:2301.07443 [hep-ph]].

\bibitem{Wang:2019dls}
D.~Wang,
Eur. Phys. J. C \textbf{79}, no.5, 429 (2019)
[arXiv:1901.01776 [hep-ph]].

\bibitem{Zhang:2025jnw}
B.~n.~Zhang and D.~Wang,
Phys. Lett. B \textbf{868}, 139674 (2025)
[arXiv:2503.21885 [hep-ph]].

\bibitem{Wang:2021rhd}
D.~Wang,
JHEP \textbf{03}, 155 (2022)
[arXiv:2111.11201 [hep-ph]].

\bibitem{Wang:2022wrb}
D.~Wang,
Phys. Rev. D \textbf{105}, no.7, 073002 (2022)
[arXiv:2203.02930 [hep-ph]].

\bibitem{tHooft:1973alw}
G.~'t Hooft,
Nucl. Phys. B \textbf{72}, 461 (1974).

\end{thebibliography}
\end{document}